\documentclass[fleqn,usenatbib]{mnras}

\usepackage{newtxtext,newtxmath}

\usepackage[T1]{fontenc}
\usepackage[dvipsnames]{xcolor}
\usepackage[normalem]{ulem}

\DeclareRobustCommand{\VAN}[3]{#2}
\let\VANthebibliography\thebibliography
\def\thebibliography{\DeclareRobustCommand{\VAN}[3]{##3}\VANthebibliography}

\usepackage{graphicx}	
\usepackage{amsmath}	

\title[Magnetic field effects on HMNS systems]{
Magnetic field effects on nucleosynthesis and kilonovae from neutron star merger remnants \\
} 

\author[S. E. M. de Haas]{
Sebastiaan de Haas$^{1}$\thanks{E-mail: sebastiaan.dehaas@student.uva.nl},
Pablo Bosch$^{1}$,
Philipp M{\"o}sta$^{1}$,
Sanjana Curtis$^{2,3}$
and Nathanyel Schut$^{1}$
\\
$^{1}$Gravitation Astroparticle Physics Amsterdam (GRAPPA) Institute, University of Amsterdam,
Science Park 904, 1098 XH Amsterdam, The Netherlands\\
$^{2}$Department of Astronomy \& Astrophysics, University of Chicago, 5640 S Ellis Avenue, Chicago, IL 60637, USA\\
$^{3}$Kavli Institute for Cosmological Physics, University of Chicago, Chicago, IL 60637, USA\\
}

\date{Accepted XXX. Received YYY; in original form ZZZ}

\pubyear{2022}

\begin{document}
\label{firstpage}
\pagerange{\pageref{firstpage}--\pageref{lastpage}}
\maketitle

\begin{abstract}

We investigate the influence of parametric magnetic
field configurations of a hypermassive neutron star (HMNS) on electromagnetic
(EM) observables, specifically the kilonova lightcurves and nucleosynthesis yields.
We perform three-dimensional (3D) dynamical-spacetime general-relativistic
magnetohydrodynamic (GRMHD) simulations, including a neutrino leakage scheme, microphysical
finite-temperature equation of state (EOS), and an initial poloidal magnetic
field. We find that varying the magnetic field strength and falloff impacts
the formation of magnetized winds or mildy-relativistic jets, which in turn has
profound effects on the outflow properties. All of the evolved configurations collapse to a black hole (BH) $\sim 21-23$ ms after the onset of the simulations, however, the ones
forming jets may be considerably more effective at transporting angular
momentum out of the system, resulting in earlier collapse times. Larger mass ejecta rates and radial
velocities of unbound material characterise the systems that form jets. 
The bolometric light curves of the kilonovae and $r$-process yields change considerably with different magnetic field parameters. We conclude that 
the magnetic field strength and falloff have robust effects on the outflow 
properties and electromagnetic observables. This can be particularly important as the total ejecta mass from our simulations ($\simeq 10^{-3}\;M_{\odot}$) makes the ejecta 
from HMNS a compelling source to power kilonova through radioactive decay
of $r$-process elements.

\end{abstract}


\begin{keywords}
stars: magnetars -- (magnetohydrodynamics) MHD -- methods: numerical -- nuclear reactions, nucleosynthesis, abundances
\end{keywords}



\section{Introduction}

Multi-messenger observations of GW170817 have confirmed that binary neutron
star (BNS)  merger remnants can launch short gamma-ray bursts \citep[sGRB, e.g.,][]{2017ApJ...848L..12A,2017ApJ...848L..13A,2017ApJ...848L..15S}. Moreover,
the UV, optical and (near-)infrared observations of the BNS merger show that the
radioactive decay of rapid-neutron capture process ($r$-process) elements is
taking place in the ejecta. \citep[e.g.,][]{2017Sci...358.1574S,2017Natur.551...75S,2017ApJ...848L..19C,2017Natur.551...67P}.
Different engine models have been proposed, however late-time kilonova 
emission and sGRB observations have placed constraints on their characterization; 
a consensus on whether the remnant was a black hole or a magnetar is yet to be 
reached \citep[e.g.,][]{2017ApJ...850L..19M,2017PhRvD..96l3012S,2018ApJ...856..101M}. \citet{2020ApJ...901L..37M}
showed that magnetars formed in BNS mergers are, 
indeed, viable 
candidates for powering sGRB.

$r$-process nucleosynthesis in the BNS merger ejecta
produces large amounts of radioactive material, powering kilonova transients
while producing the heaviest elements in the universe
\citep[e.g.,][]{2011ApJ...738L..32G,2017Natur.551...80K,2021RvMP...93a5002C}.
The extensively studied kilonova related to GW170817, AT2017gfo, displayed a two-component emission.  
The ``blue'' component is associated to the early phase of the 
BNS merger with an emission peak in the UV/optical bands, while 
the ``red'' component peaks in the (near-)infrared
frequencies 
on the order of a few days post-merger
\citep[e.g.,][]{2017Sci...358.1574S,2017Natur.551...75S}. The blue component
is thought to arise from lanthanide- and neutron-poor ejecta with the majority
of emission originating from light elements \citep[with atomic number $A < 140$
and particularly large amounts of iron,
e.g.,][]{2017ApJ...848L..19C,2017ApJ...848L..18N}.  The red component would
then be dominated by emission from heavily synthesized material as a result of
$r$-process nucleosynthesis (nuclei with $A > 140$), therefore being
lanthanide- and neutron-rich
\citep[e.g.,][]{2017Natur.551...67P,2017ApJ...848L..27T,2017ApJ...848L..19C}.
Furthermore, analysis of a large electromagnetic (EM) data set conducted by
\citet{2017ApJ...851L..21V} implied that for the red component, a delayed
outflow  from the remnant accretion disk is the most likely dominant origin of
emission, in combination with an emission component from the dynamical ejecta.
The origin of the blue component is not as well understood, as it has proven
difficult to reproduce the inferred outflow properties with simulations
\citep{2018ApJ...869L...3F}. Among the suggested possibilities are shock-heated
polar dynamical ejecta \citep[e.g.,][]{2017arXiv171005931M}, neutrino-driven  winds from the HMNS remnant,
magnetized winds from the HMNS remnant \citep[see also][]{2018ApJ...856..101M}
and remnant winds from spiral density waves \citep{2019ApJ...886L..30N}, where the final two seem the most
promising. Furthermore, the EM data analysed by \citet{2017ApJ...851L..21V} implies a
blue kilonova component with an ejecta mass $M_{\rm ejecta}$ of $\approx 2.0
\times 10^{-2} M_{\odot}$ and ejecta speed $v_{\rm ejecta} \approx 0.27c$ and
a red component with $M_{\rm ejecta} \approx 1.1 \times 10^{-2} M_{\odot}$ and
$v_{\rm ejecta} \approx 0.14c$.

BNS post-merger remnants may be highly magnetized following an amplification stage as a
result of magnetic instabilities, such as the Kelvin-Helmholtz instability in
the shear layer between two streams of matter during the pre-merger phase
\citep[e.g.,][]{2013ApJ...769L..29Z,2015PhRvD..92l4034K}. The strong magnetic
field that is generated ,likely, has profound effects on the remnant system.
Therefore, simulations of BNS mergers increasingly account for magnetic field
effects by implementing general-relativistic magnetohydronamic (GRMHD) methods
\citep[e.g.,][]{2009MNRAS.399L.164G,2015PhRvD..92l4034K,2015PhRvD..92h4064D,2019PhRvD.100b3005C}.
Comparisons between GRMHD and purely GRHD simulations of BNS mergers have
implied robust effects of the magnetic field on outflow properties
\citep[e.g.,][]{2008PhRvL.100s1101A,2008PhRvD..78b4012L,2018PhRvD..97l4039K}.
Namely, it may cause the formation of mildly-relativistic jets and results in
considerably larger mass ejecta rates and ejecta velocities \citep{2020ApJ...901L..37M}.

As GRMHD and GRHD simulations of BNS mergers imply strong magnetic field
effects on outflow properties, it is interesting to parametrically explore the
influence of the magnetic field by varying its strength and configuration. 
\citet{2014ApJ...785L...6S} investigated the latter, in the context of BNS merger remnants, using three different magnetic field geometries to determine their influence on the X-ray afterglow of the sGRB. They evolved an initially isolated axisymmetric HMNS, with a polytropic equation of state and endowed with a magnetic field, rather than the direct outcome of a BNS merger evolution.
In this work, we perform seven dynamical-spacetime GRMHD simulations of (post-merger) 
hypermassive neutron star (HMNS) systems including a parameterized magnetic 
field with different field strengths and configurations, to investigate 
the influence of these magnetic field parameters on the HMNS outflows and kilonova. 
We map a snapshot
of BNS post-merger data, at $t_{\rm map} = 17$ ms after coalescence, from a
GRHD simulation performed by \citet{2018ApJ...869..130R} and use it as initial
data for all the simulations. We post-process the HMNS ejecta, using Lagrangian
tracer particles, to compute the $r$-process yields and
a spherically-symmetric radiation-hydrodynamics code to compute bolometric
light curves of the kilnovae. Both magnetic field parameters show profound effects on the computed 
outflow properties, nucleosynthesis yields and kilonova light curves. All simulations
collapse to a BH $\sim 38 - 40$ ms after coalescence of the two neutron stars. 
Two of the seven simulations show the emergence of
mildly-relativistic jets, while displaying significantly earlier BH collapse
times compared to the other simulations (by $\sim 1.6$ ms). This may imply that
jets are more effective at transporting angular momentum out of the remnant
system compared to magnetized winds. Furthermore, the two simulations that
exhibit jet formation contain significantly larger mass ejecta rates and
radial velocities of unbound material. We find that the total ejecta mass of the 
HMNS system is in the $2.4 \times 10^{-4}\,M_{\odot} < M_{\rm ejecta} < 8.3 \times
10^{-3}\,M_{\odot}$ range for all seven  simulations. Finally, we show that the magnetic field has significant implications on the nucleosynthesis yields and kilonova light curves even for the weaker magnetic field range explored, thus making this a robust feature for magnetized HMNS remnants.

The paper is organized as follows. In Section \ref{sec:methods}, we describe
our simulation setup, numerical methods and the procedure for obtaining the
$r$-process yields and kilonova light curves.
In Section~\ref{sec:outflowproperties}, we discuss the various black hole collapse times
and outflow properties of the HMNS system, followed by the evolution of the
magnetic vector field in Section \ref{sec:magnetic-properties}. We discuss the
nucleosynthesis yields and bolometric light curves of the kilonovae in Section
\ref{sec:nucleo-and-kilonovae}. We summarize and discuss our conclusions in
Section \ref{ch:summary-conclusions}.

\section{Numerical Methods and Setup}
\label{sec:methods}

The simulations performed in this work make use of the Einstein toolkit framework
\citep{2012CQGra..29k5001L}, which is a publicly-available infrastructure for relativistic astrophysics and gravitational physics simulations (\url{http://einsteintoolkit.org}). 
The code is based on multiple components, including the \texttt{Carpet} thorn that is
responsible for adaptive mesh refinement (AMR) \citep{2004CQGra..21.1465S}, the
code that provides GRMHD named \texttt{GRHydro} \citep{2014CQGra..31a5005M} and
the \texttt{McLachlan} module that generates GR evolution
\citep{2009PhRvD..79d4023B,2011PhRvD..83f4008R}. We use finite-volume high-resolution
shock capturing (HRSC) methods to evolve the system in time and adopt 5th-order 
weighted-ENO (WENO5) reconstruction
\citep{2007MNRAS.379..469T,2013PhRvD..87f4023R}
and the HLLE approximate Riemann solver
\citep{1983JCoPh..49..357H,1988SJNA...25..294E}. To prevent violations of the magnetic field divergence-free constraint, $\vec{\nabla} \cdot \vec{B} = 0$, we enforce them through a constrained transport scheme.

\subsection{Equation of state and neutrino treatment}

For the simulations performed in this work we adopt a microphysical, finite-temperature
equation of state (EOS) in tabulated form. Specifically, we use the $K_0 = 220$ MeV
variant of the EOS from \citet{1991NuPhA.535..331L} (where $K_0$ is the nuclear compression modulus), which is the
so-called LS220 EOS.

The simulations include a neutrino treatment through a scheme that adopts
neutrino heating and leakage approximations, based on
\citet{2010CQGra..27k4103O} and \citet{2013ApJ...768..115O} (which in turn are
based on \citet{2003MNRAS.342..673R} and \citet{1996A&A...311..532R}). The
scheme tracks three different neutrino species; electron neutrinos $\nu_e$,
electron anti-neutrinos $\bar{\nu}_e$ and heavy-lepton tau and muon
(anti-)neutrino's, which are grouped in a single neutrino species $\nu_x = \{
\nu_{\mu}, \nu_{\tau}, \bar{\nu}_{\mu}, \bar{\nu_{\tau}} \}$. This grouping is
reasonable as these neutrinos interact only through neutral current processes in
the post-merger environment, which contain similar cross sections. The following
interactions are included in estimates for the neutrino energy and number
emission rates; the charged-current capture processes

\begin{equation}
    p + e^{-} \leftrightarrow n + \nu_e\:,
    \label{eq:electronneutrino}
\end{equation}
\vspace{-.5cm}
\begin{equation}
    n + e^{+} \leftrightarrow p + \bar{\nu}_e\:,
    \label{eq:electronantineutrino}
\end{equation}

plasmon decay,

\begin{equation}
    \gamma \leftrightarrow \nu + \bar{\nu}\,,
\end{equation}

electron and positron pair annihilation/creation,

\begin{equation}
    e^{-} + e^{+} \leftrightarrow \nu + \bar{\nu}\,,
\end{equation}

and nucleon-nucleon Bremsstrahlung,

\begin{equation}
    N + N \leftrightarrow N + N + \nu + \bar{\nu}\,,
\end{equation}

where the approximate neutrino energy and number emission rates from the above
processes depend on local thermodynamics and the energy-averaged optical depth.
Estimates for the neutrino optical depth are based on non-local calculations,
which have been implemented using a ray-by-ray approach. The scheme solves the
neutrino optical depth along radial rays that cover the simulation domain using
the $\theta$ and $\psi$ directions. Tri-linear interpolation is then used in
spherical coordinates $(r,\theta,\psi)$ for determining the optical depth at
Cartesian grid cell centers. For the simulations, 20 rays in $\theta$ are
employed that cover $[0,\pi/2]$ and 40 rays in $\psi$, covering $[0,2\pi]$. The
rays contain 800 equidistant points each up to a distance of 120 km, after
which 200 logarithmically spaced points are adopted to account for the
remainder of the domain.

The approximated local neutrino heating function is based on the charged-current absorption of $\nu_e$ and $\bar{\nu}_e$, \eqref{eq:electronneutrino} and \eqref{eq:electronantineutrino}, and is given by

\begin{equation}
    \mathcal{Q}^{\rm heat}_{\nu_i} = f_{\rm heat} \frac{L_{\nu_i}(r)}{4 \pi r^2} \langle \epsilon^2_{\nu_i} \rangle S_{\nu} \frac{\rho}{m_n} X_i \biggl< \frac{1}{F_{\nu_i}} \biggr> e^{-2 \tau_{\nu_i}}\,,
    \label{eq:neutrinoheating}
\end{equation}

 where $f_{\rm heat}$ is the heating scale factor, $L_{\nu_i}(r)$ the
approximate neutrino luminosity that emerges radially from below as
interpolated by the ray-by-ray approach of the neutrino leakage scheme and $~{S_{\nu} =0.25 \ (1 + 3 \alpha^2)\ \sigma_0\ \frac{1}{(m_e c^2)}}$, where $\alpha = 1.23$,
$\sigma_0 = 1.76 \times 10^{-44}\,\mathrm{cm}^{-2}$, $m_e$ the electron mass and $c$
the speed of light. Additionally, $\langle \epsilon^2_{\nu_i} \rangle$ is the approximate
neutrino mean-squared energy, $m_n$ the neutron mass, $X_i$ is the neutron or
proton mass fraction for the electron neutrino's or anti-neutrinos,
respectively, $\langle \frac{1}{F_{\nu_i}} \rangle$ is the mean inverse flux
factor and $\tau_{\nu_i}$ is the approximate neutrino optical depth. More
specifically, $\langle \frac{1}{F_{\nu_i}} \rangle$ depends on neutrino
radiation field details and is parameterized as a function of $\tau_{\nu_i}$,
based on neutrino transport calculations from \cite{2008ApJ...685.1069O} and
given by $\langle \frac{1}{F_{\nu_i}} \rangle = 4.275 \tau_{\nu_i} + 1.15$.
Furthermore, the heating scale factor $f_{\rm heat}$ is a free parameter that
has been set to $f_{\rm heat} = 1.05$, which is consistent with heating in
core-collapse supernova simulations that adopt full neutrino transport schemes
\citep{2013ApJ...768..115O}. The above neutrino heating function was first
derived by \citet{2001A&A...368..527J}. Neutrino heating is turned off in
the simulations for densities $\rho < 6.18 \times 10^{10}$ g cm$^{-3}$, in
order to maintain numerical stability. The neutrino scheme correctly captures
the overall neutrino energetics up to a factor of a few when compared to the
full neutrino transport scheme of \citet{2010CQGra..27k4103O} for simulations
of core-collapse supernovae. The dependence on energy, the deposition of
momentum and the annihilation of neutrino pairs are not included in the scheme,
and consequently will likely affect our inferred composition properties of the
ejecta.

\subsection{Initial conditions of the simulations}

The initial data is mapped from a GRHD simulation of a BNS merger by
\citet{2018ApJ...869..130R}, covering both the pre-merger phase and a small
fraction of the post-merger phase. This simulation is based on the
\texttt{WhiskyTHC} code (model LS135135M0), and evolves an equal-mass binary NS with component masses at infinity of $1.35 M_{\odot}$, the same EOS
and similar neutrino treatment. The mapping of this simulation is done at a
time $t_{\rm map} - t_{\rm merger} = 17$ ms, thereby avoiding transient,
oscillatory effects caused by the NS remnant core in the early post-merger
phase. 

Five different AMR levels are implemented, varying by a factor of two in resolution between consecutive levels. The highest refinement level region, covering the HMNS, has a resolution
$h_{\rm fine} = 185$ m, while for the coarsest region $h_{\rm coarse} = 3.55$
km. The structure of the AMR grid is made up of boxes that extend up
to 177.3km, 118.2 km, 59.1 km and 29.6 km, while the outermost boundary of the
simulation domain extends to a distance of $\sim 355$ km. 

At the onset of our HMNS simulations, we add a parameterized magnetic field to the simulations, which varies in strength and falloff between the different simulations. We initialize the parameterized magnetic field with the analytical prescription of the vector potential $\vec{A}$, where $\vec{B} = \nabla \times \vec{A}$, of the form

\begin{equation}
    A_r = A_\theta = 0; \quad A_\phi = B_0\,r\,\sin(\theta) \frac{r_{\rm falloff}^3}{r_{\rm falloff}^3 + r^3}\:,
\label{eq:initialBfield}
\end{equation}

where $B_0$ is the initial magnetic field strength and $r_{\rm falloff}$
controls the range of the magnetic field. As we add this purely poloidal, large-scale magnetic field \emph{ad hoc}, we implicitly 
assume that a dynamo process is present during the pre-merger (and possibly
also early post-merger) phase that is capable of producing such an ordered,
strong field. Even though previous research of proto-neutron stars formed in core-collapse supernovae implies the presence of such a
dynamo \citep[e.g.,][]{2015Natur.528..376M,2020SciA....6.2732R}, current
BNS merger simulations are not capable of fully resolving this
magnetic amplification process \citep[e.g.,][]{2018PhRvD..97l4039K}.

We perform a total of seven simulations. For the first three simulations, we
vary the magnetic field strength between $B_0 = \{10^{13},\, 10^{14},\, 10^{15}\}$ G while keeping the magnetic falloff parameter $r_{\rm falloff} = 20$
km fixed. For the next three simulations, we fix the magnetic field strength $B_0 = 10^{15}$ G while varying
$r_{\rm falloff}$ between $r_{\rm falloff} = \{5,\, 10,\,15\}$ km. For the final
simulation, we change both magnetic field parameters, explicitly, $B_0 = 5 \times
10^{15}$ G and $r_{\rm falloff} = 10$ km. We list the values of the magnetic
field parameters of the seven simulations in Table
\ref{table:sevensimulations}, and include corresponding nomenclature for the simulations. 

\begin{table}
\centering
\begin{tabular}{ | c | c | c | }
Simulation name & $B_0$ [Gauss] & $r_{\rm falloff}$ [km] \\
\hline
B15-r20 & $10^{15}$ & 20 \\
B14-r20 & $10^{14}$ & 20 \\
B13-r20  & $10^{13}$ & 20 \\
B15-r5  & $10^{15}$ & 5 \\ 
B15-r10  & $10^{15}$ & 10 \\
B15-r15  & $10^{15}$ & 15 \\
B5-15-r10 & $5 \times 10^{15}$ & 10 
\end{tabular}
\caption{\label{table:sevensimulations} Initial conditions of the various magnetic fields that have been adopted during the seven performed simulations of this work. The parameter $B_0$ controls the magnetic field strength, while $r_{\rm falloff}$ is responsible for the range the magnetic field. For the mathematical form of the vector potential of the magnetic field, see equation \eqref{eq:initialBfield}.}
\end{table}

\subsection{Nucleosynthesis and kilonova analysis}
\label{sec:nucleo-analysis}

To calculate the nucleosynthesis yields, we use Lagrangian tracer particles to 
determine the encountered neutrino luminosities and thermodynamic 
quantities of the merger outflows. The tracer particles are spaced uniformly 
and we extract the corresponding quantities once the tracers reach a distance 
of $r = 150\: M_{\odot}$. We determine the composition of the merger ejecta 
by post-processing the tracers using the nuclear reaction network \texttt{SkyNet} 
\citep{2017ApJS..233...18L}. REACLIB is used to obtain the forward strong 
rates, nuclear masses, partition functions and part of the weak rates 
\citep{2010ApJS..189..240C}. The remaining weak rates are taken from 
\citet{1982ApJS...48..279F}, \citet{1994ADNDT..56..231O} or \citet{2000NuPhA.673..481L}. 

Note that we adopt an approximate neutrino leakage 
scheme in the simulations, while the ejecta composition depends sensitively 
on the neutrino transport performed by this scheme. This causes uncertainties in 
our predictions of $Y_e$ distributions and $r$-process abundances. These uncertainties have been investigated by \citet{2021arXiv211200772C}, where various neutrino luminosities have been adopted to determine its influence on the $r$-process abundances and $Y_e$ distributions. They conclude that the $r$-process production of heavy elements is reduced by up to a factor of $\sim$10 when comparing the
two most extreme cases that bracket the entire adopted parameter space.

In order to compute the luminosity of the kilonova on a timescale of days, 
we use a modification of \texttt{SNEC} \citep[SuperNova Explosion Code;][]{2015ApJ...814...63M}, 
which is a 1D Lagrangian equilibrium-diffusion radiation hydrodynamics code 
that can simulate the evolution of merger outflows and consequent kilonova 
emission. Modifications to \texttt{SNEC} are implemented to account for kilonova 
as opposed to supernova modeling, such as the nickel heating term which is 
replaced by radioactive heating from $r$-process nuclei. We follow the same procedure 
as \citet{2021arXiv211200772C}, where more details on 
the modifications and 
methods of the kilonova modeling and on the post-processed 
nucleosynthesis can be found.

\section{Results}
\label{sec:results}

\subsection{Black hole collapse and outflow properties}
\label{sec:outflowproperties}

\begin{figure}
    \centering
    \includegraphics[scale=0.33]{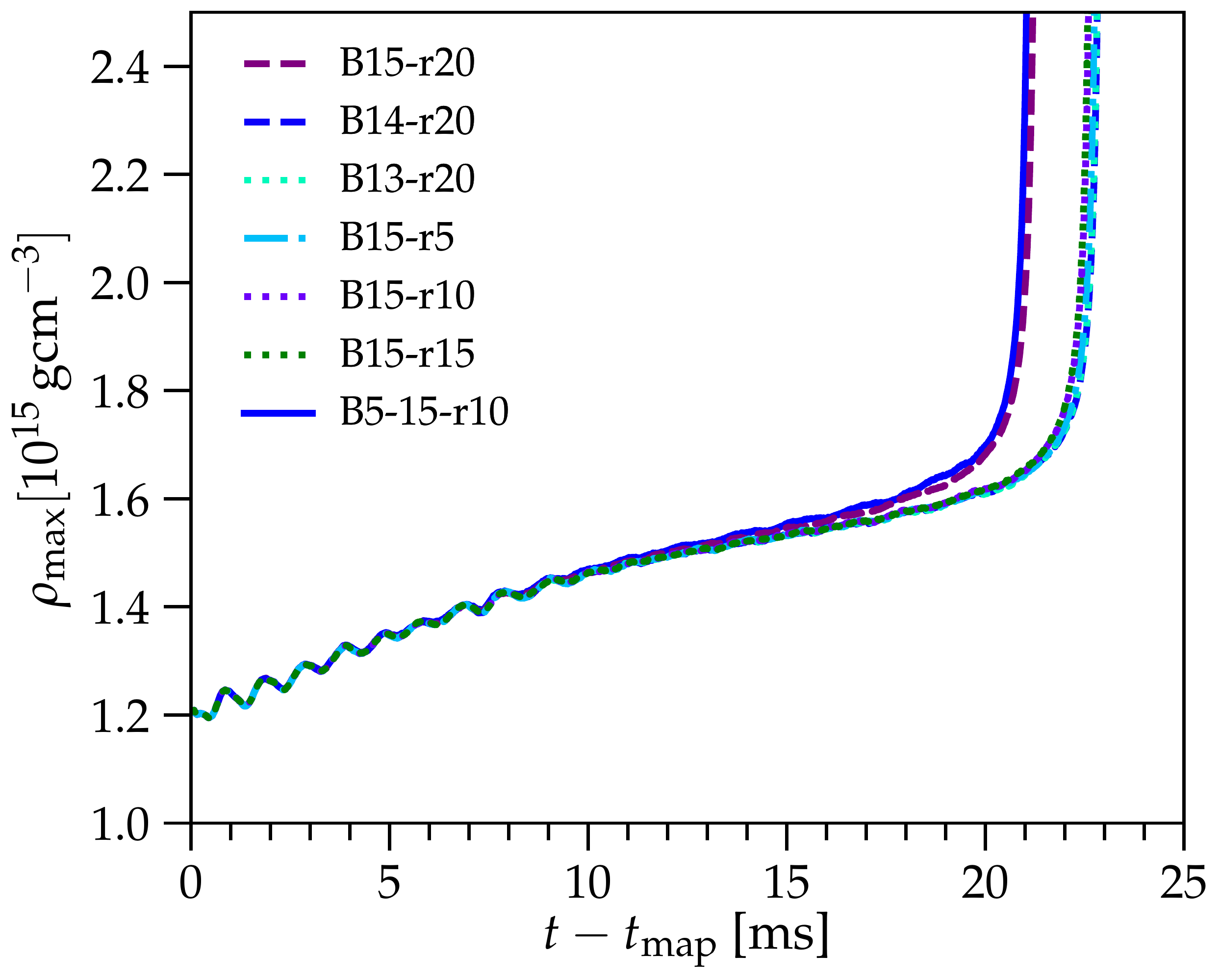}
    \caption{The maximum density $\rho_{\rm max}$ as a function of time for all simulations. Simulations B15-r20 and B5-15-r10 display earlier BH collapse times of $\sim 21.3$ ms compared to the other simulations, which collapse after $\sim 22.9$ ms.}
    \label{fig:rhomaxvstime}
\end{figure}

In Fig. \ref{fig:rhomaxvstime}, we show the maximum density $\rho_{\rm max}$ as a function of time for all simulations. Simulations B15-r20 and B5-15-r10 collapse to a BH after $\sim 21.3$ ms, while the other simulations show an on-average increased collapse time of $\sim 1.6$ ms at $\sim 22.9$ ms (all simulations display slight differences in the exact collapse times). The significant difference in collapse time of $\sim 1.6$ ms between these two groups of simulations may be explained by the formation of collimated, mildly-relativistic jets for these two simulations. Even though all simulations launch magnetized winds along the rotation axis of the HMNS remnant \citep{2004ApJ...611..380T}, only for the two aforementioned simulations is the magnetic field powerful enough to collimate part of the outflow from the HMNS into jets.

\begin{figure*} 
\centering
	\begin{minipage}{0.3\textwidth}
	 \includegraphics[width=\textwidth]{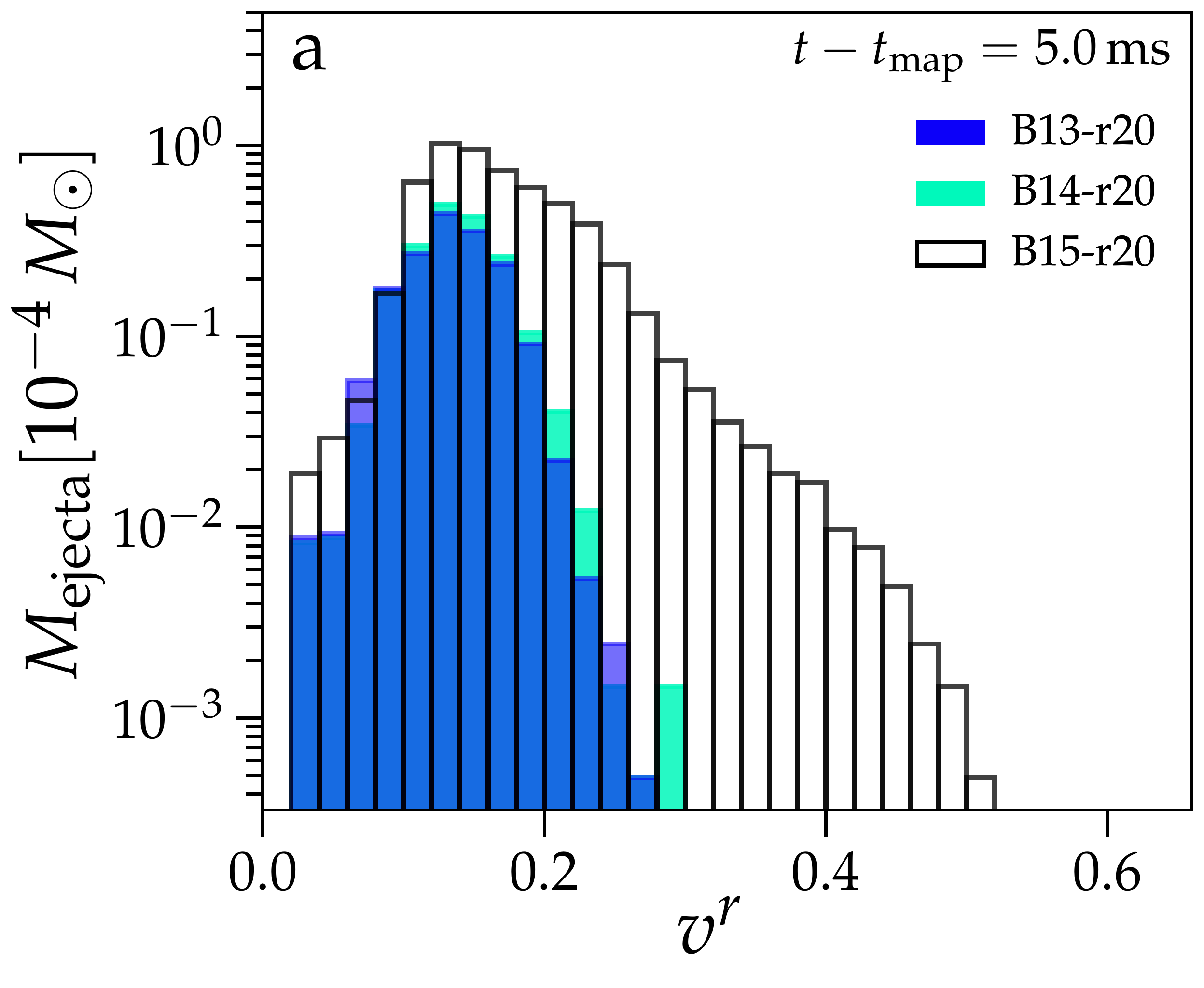}
	\end{minipage}
  \hfill 
  \begin{minipage}{0.3\textwidth}
   \includegraphics[width=\textwidth]{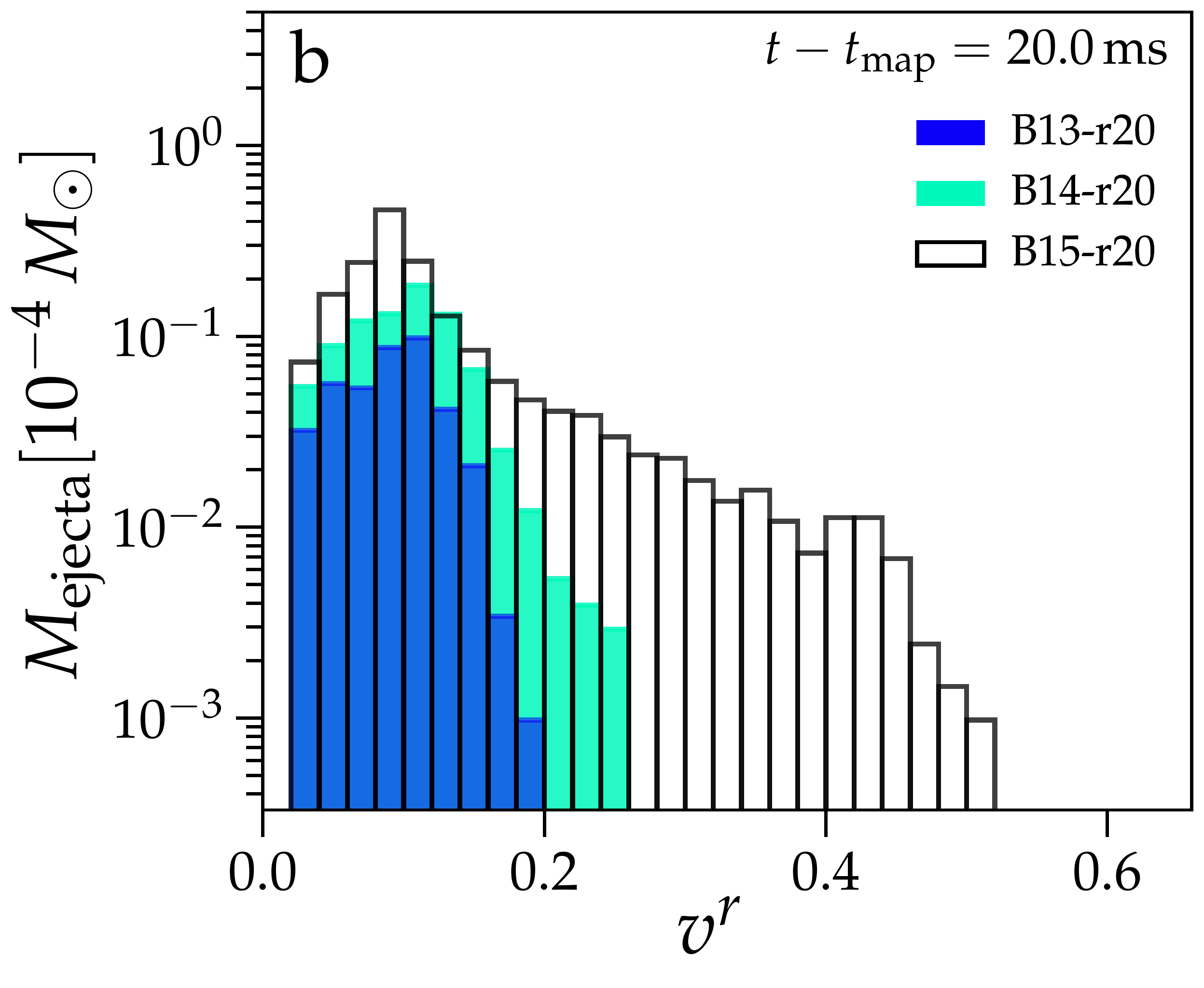}
  \end{minipage}
  \hfill
  \begin{minipage}{0.3\textwidth}
   \includegraphics[width=\textwidth]{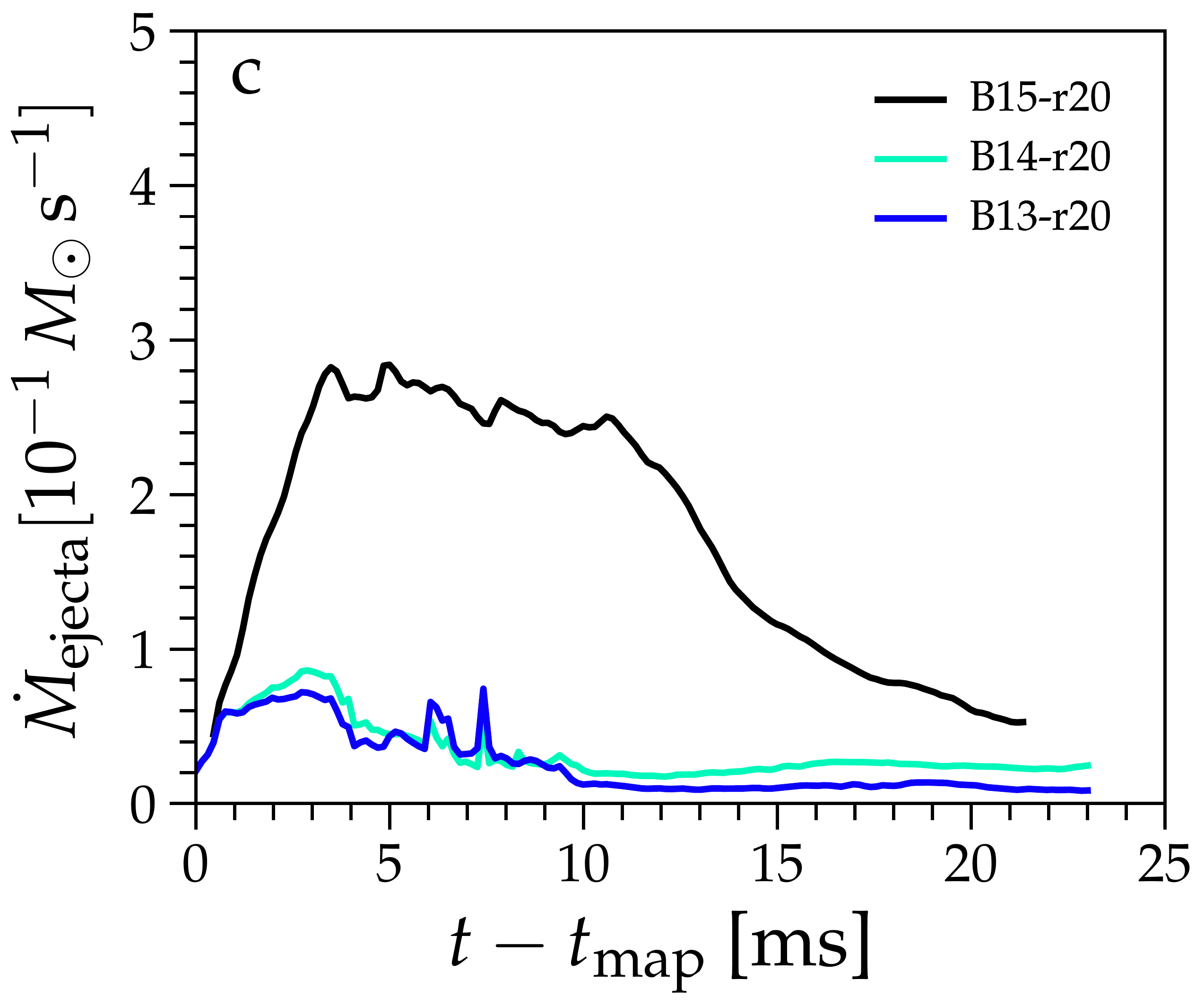}
  \end{minipage}\\
	\begin{minipage}{0.3\textwidth}
	 \includegraphics[width=\textwidth]{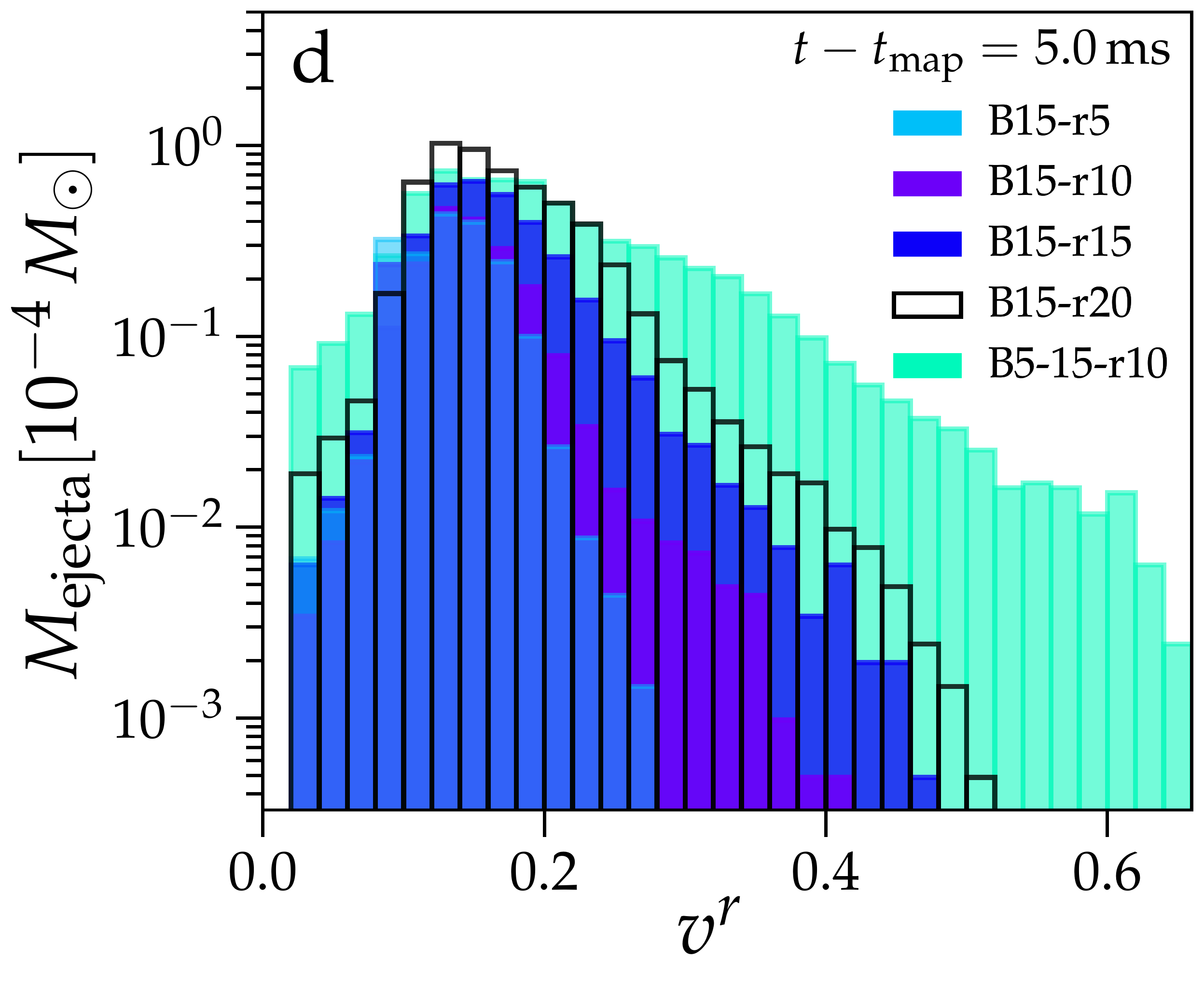}
	\end{minipage}
  \hfill 
  \begin{minipage}{0.3\textwidth}
   \includegraphics[width=\textwidth]{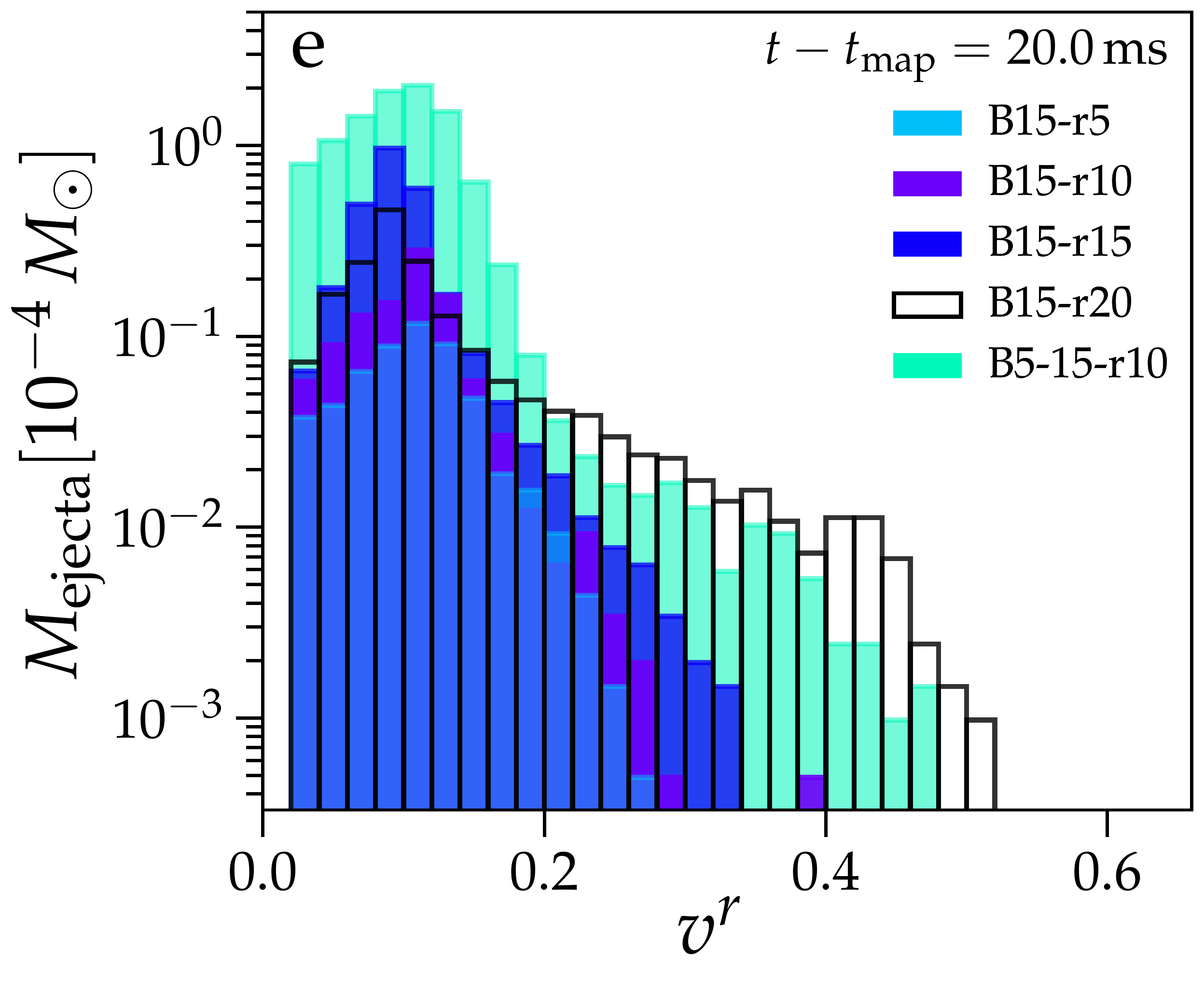}
  \end{minipage}
  \hfill
  \begin{minipage}{0.3\textwidth}
   \includegraphics[width=\textwidth]{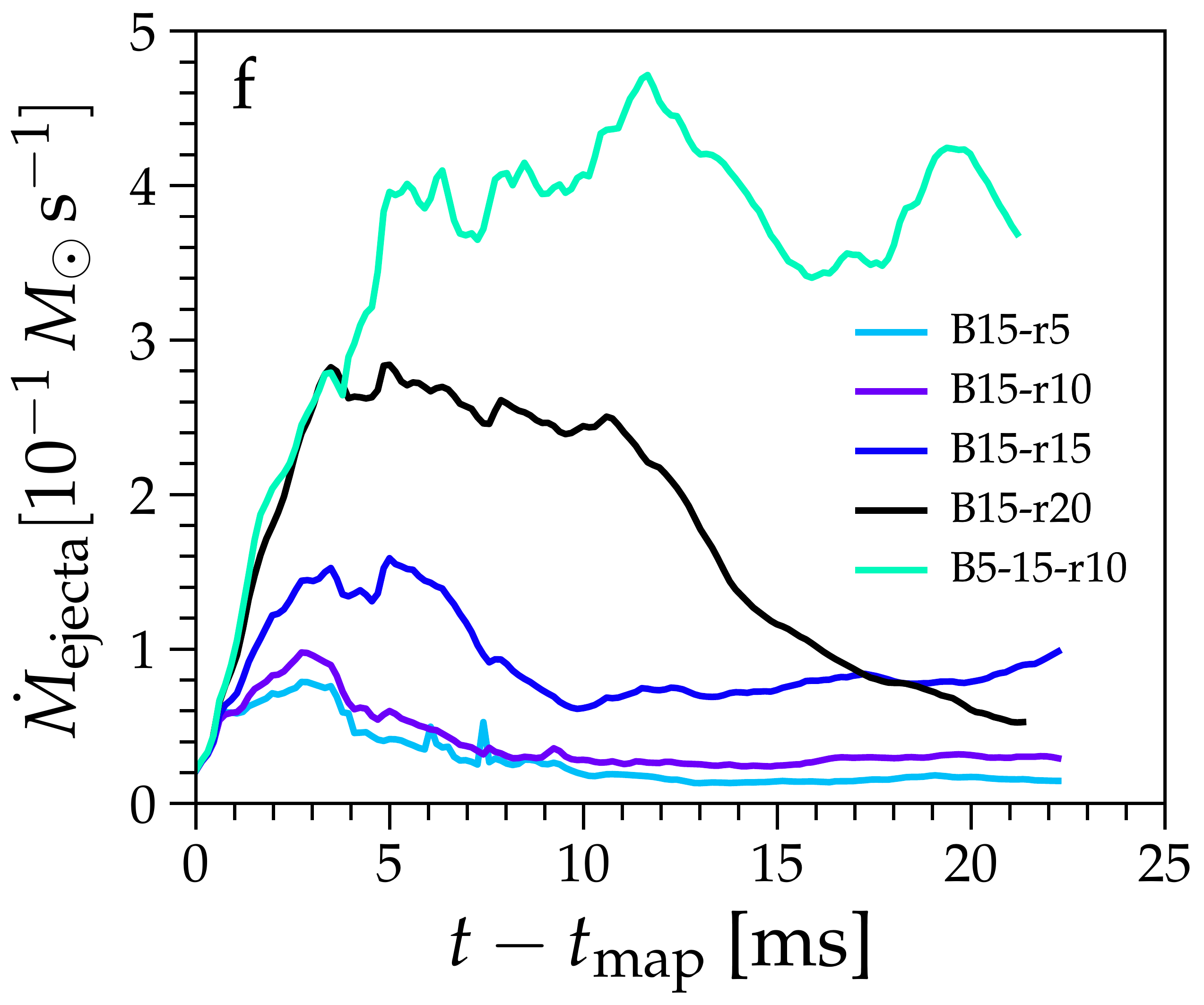}
  \end{minipage}
    \caption{\textbf{Panels a, b and c:} Comparison between simulations with various initial magnetic field strengths $B_0$, where we show B13-r20 in blue, B14-r20 in green and B15-r20 in black. In panels a and b, we show histograms of the radial velocity $v^r$ of unbound material (defined as material satisfying the Bernoulli criterion $-h u_t > 1$) with corresponding ejecta masses $M_{\rm ejecta}$ at $t - t_{\rm map} = 5$ and 20 ms. In panel c, we show the mass ejecta rate $\dot{M}_{\rm ejecta}$ of unbound material as a function of time. \textbf{Panels d, e and f:} Comparison between simulations with different magnetic fall-off parameter $r_{\rm falloff}$ (where B5-15-r10 is also included), where we show B15-r5 in cyan, B15-r10 in purple, B15-r15 in blue, B15-r20 in black and B5-15-r10 in green. We show histograms of the radial velocity $v^r$ of unbound material at $t - t_{\rm map} = 5$ and 20 ms in panels d and e and the mass ejecta rate $\dot{M}_{\rm ejecta}$ of unbound material as a function of time in panel f.}
	\label{fig:Vr-histograms-mdot} 
\vspace{0.5cm} 
\end{figure*}

In order to evaluate the properties of unbound material exclusively, we calculate the material's Bernoulli criterion $-h u_t > 1$, where $h = (1 + \epsilon + p + \frac{b^2}{2})/\rho$ is the fluid's relativistic enthalpy and $u_t$ the time component of the fluid four-velocity. If the Bernoulli criterion is satisfied, the corresponding material is unbound. In the upper row of Fig. \ref{fig:Vr-histograms-mdot}, we show histograms of the velocity's radial component $v^r$ of unbound material with corresponding ejecta mass $M_{\rm ejecta}$ for simulations B13-r20, B14-r20 and B15-r20 at $t - t_{\rm map} = 5$ and 20 ms. In addition, we show the evolution of the sphere-averaged mass ejecta rates $\dot{M}_{\rm ejecta}$ as a function of time for the same simulations, which are computed using

\begin{equation}
    \dot{M}_{\rm ejecta} = \int_{r_1}^{r_2} \sqrt{g} \rho W v^r dV \frac{1}{(r_2 - r_1)}\:,
\end{equation}

with $r_1 = 44.3$ km and $r_2 = 192.1$ km. Material is only included in this computation if the Bernoulli criterion is satisfied. We show B15-r20 \citep[which is almost identical to B15-low in][]{2020ApJ...901L..37M} as a reference case in black\footnote{We modified how tracer particles record neutrino luminosities in low-density regions.}.

For the $v^r$ evolution at $t - t_{\rm map} = 5$ ms, B13-r20 and B14-r20 display very similar $v^r$ profiles with $v^r < 0.3c$. Simulation B15-r20 contains significantly larger ejecta masses for nearly all $v^r$, while also displaying ejecta in the $0.3c < v^r < 0.5c$ regime. By $t - t_{\rm map} = 20$ ms, the ejecta mass across all velocity bins have decreased significantly for all simulations. The $v^r$ profile of B14-r20 exhibits larger ejecta masses in the $v^r > 0.2c$ range while B13-r20 loses all of its ejecta in this velocity regime. For B15-r20, the mass ejecta peak has shifted to significantly lower velocities ($v^r \simeq 0.08c$).

Simulation B15-r20 shows considerably larger $\dot{M}_{\rm ejecta}$ during its evolution compared to B14-r20 and B13-r20. Simulations B14-r20 and B13-r20 exhibit very similar $\dot{M}_{\rm ejecta}$ patterns, while also displaying two short peaks at $t - t_{\rm map} \sim 6$ ms and $t - t_{\rm map} \sim 7.5$ ms. These $\dot{M}_{\rm ejecta}$ peaks are slightly enhanced for simulation B13-r20 compared to B14-r20, although the latter does generally display larger $\dot{M}_{\rm ejecta}$ values compared to the former.

In the lower row of Fig. \ref{fig:Vr-histograms-mdot}, we show $v^r$ histograms of unbound material with corresponding ejecta masses for simulations with varying $r_{\mathrm{falloff}}$, these are B15-r5, B15-r10, B15-r15, B15-r20 and B5-15-r10. At $t - t_{\rm map} = 5$ ms, all displayed simulations exhibit apparent differences in $v^r$ profiles, where especially B5-15-r10 contains large amounts of high-velocity ejecta with $0.3c < v^r < 0.66c$ while B15-r20 also shows some high-velocity outflows with $0.3c < v^r < 0.52c$. Simulations B15-r10 and B15-r15 exhibit less high-velocity ejecta with $0.3c < v^r < 0.42c$ and $0.3c < v^r < 0.48c$, respectively, while B15-r5 only contains outflows with $v^r < 0.28c$. 
At $t - t_{\rm map} = 20$ ms, the $v^r$ profiles of simulations B15-r5, B15-r10 and B15-r15 look reasonably similar, where B15-r10 and B15-r15 have lost the majority of their high-velocity ($v^r > 0.3c$) ejecta between $t - t_{\rm map} = 5$ and 20 ms. For B5-15-r10, nearly all $v^r > 0.5c$ material has rapidly decreased or disappeared in the same time interval, although it has retained significant $M_{\rm ejecta}$ values in the $0.3c < v^r < 0.5c$ regime. Simulation B15-r20, by contrast, displays larger high-velocity mass fractions at $t - t_{\rm map} = 20$ compared to $t - t_{\rm map} = 5$. Finally, we note that jet formation in simulations B15-r20 and B5-15-r10 leads to considerably larger $v^r$ values compared to their purely magnetized wind-forming counterparts.

For the corresponding $\dot{M}_{\rm ejecta}$ panel, B5-15-r10 exhibits much larger $\dot{M}_{\rm ejecta}$ values compared to the other simulations including B15-r20, despite both simulations showing jet formation. Simulation B15-r20 does exhibit significantly larger mass ejecta rates throughout most of its evolution compared to B15-r15, B15-r10 and B15-r5. Furthermore, simulation B15-r15 exhibits considerably larger $\dot{M}_{\rm ejecta}$ compared to B15-r10 and B15-r5, even showing an increasing $\dot{M}_{\rm ejecta}$ trend towards the end of the simulation. Finally, simulation B15-r5 shows a very similar $\dot{M}_{\rm ejecta}$ evolution compared to B14-r20 and B13-r20, also displaying two short peaks at $t - t_{\rm map} \sim 6$ ms and $t - t_{\rm map} \sim 7.5$ ms.

\begin{figure*}
\centering
	\begin{minipage}{0.245\textwidth}
	\includegraphics[width=\textwidth]{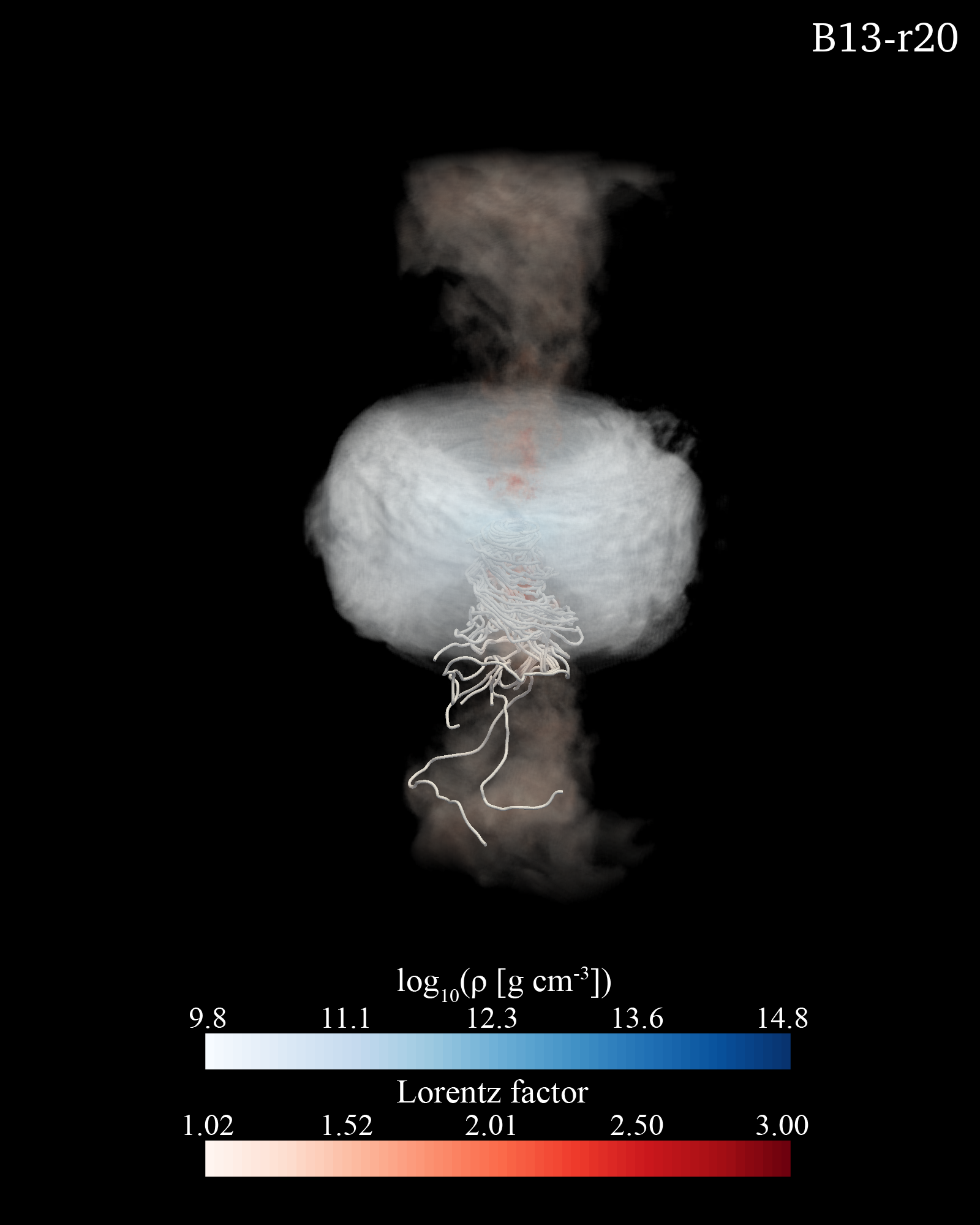}
	\end{minipage}
	\begin{minipage}{0.245\textwidth}
	\hspace{-0.15cm}
	\includegraphics[width=\textwidth]{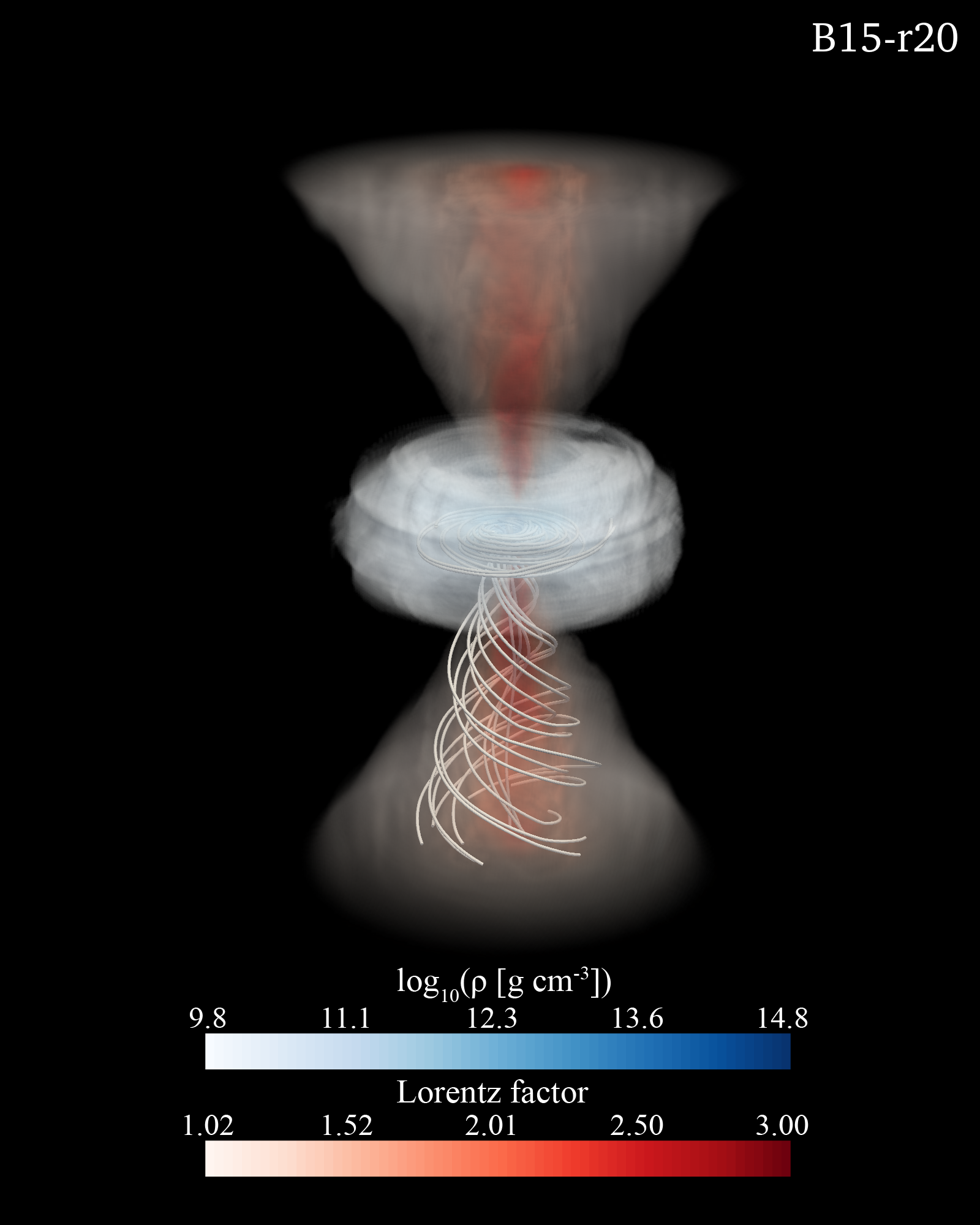}
	\end{minipage}
	\begin{minipage}{0.245\textwidth}
	\hspace{-0.3cm}
	\includegraphics[width=\textwidth]{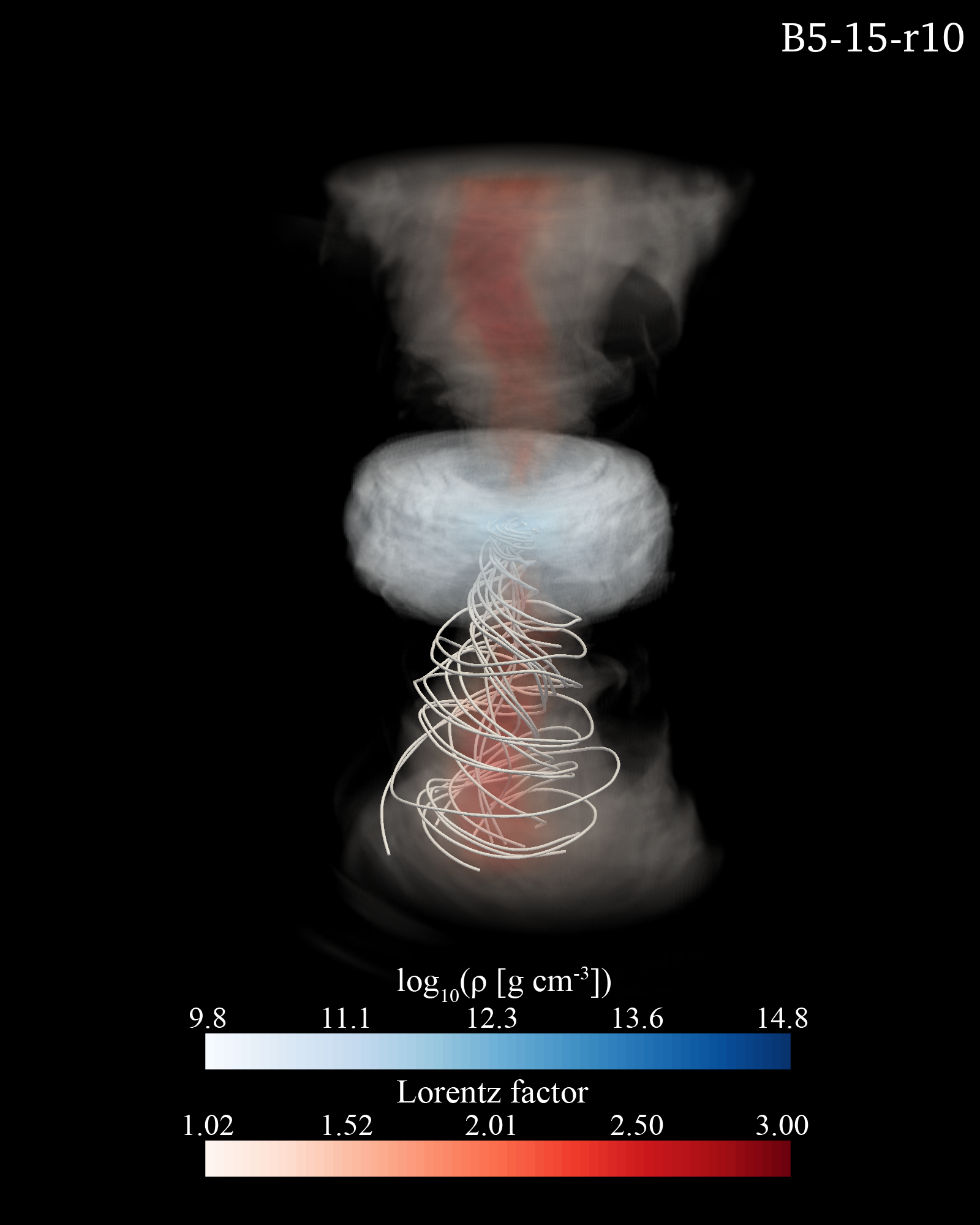}
	\end{minipage}
	\begin{minipage}{0.245\textwidth}
	\hspace{-0.45cm}
	\includegraphics[width=\textwidth]{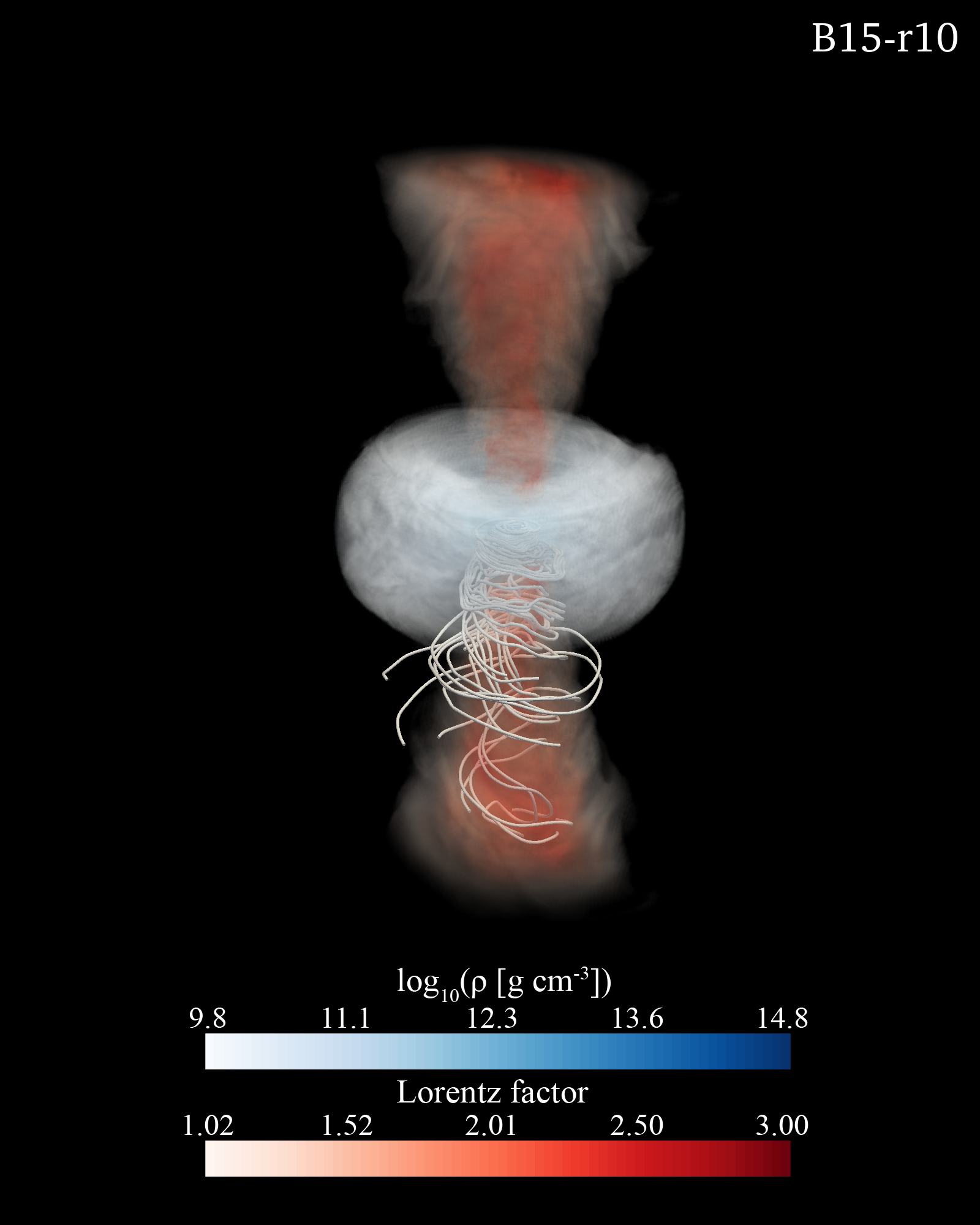}
	\end{minipage}
    \vspace{0.3cm}
	\caption{Volume renderings of the Lorentz factor (Bernoulli criterion) for the outflows (white-red colormap) and density for the accretion torus (white-blue colormap) of simulations B13-r20 (left), B15-r20 (middle-left), B5-15-r10 (middle-right) and B15-r10 (right). The magnetic field lines are also shown in the lower plane ($z < 0$, where $z$ is the vertical axis) in white. The top-to-bottom distance of the volume renderings is 355 km.}
	\label{fig:volume-renderings} 
\vspace{0.5cm} 
\end{figure*}

In Fig. \ref{fig:volume-renderings}, we show volume renderings of the Bernoulli criterion (equivalent to the Lorentz factor) for the outflows (white-red colormap) and density for the accretion torus (white-blue colormap) of simulations B13-r20, B15-r20, B5-15-r10 and B15-r10. The magnetic field lines are also shown in the lower plane ($z < 0$, where $z$ is the vertical axis) in white. When comparing B13-r20 and B15-r20, the latter shows a more structured accretion torus and a considerably larger amount of ejecta, in addition to higher Lorentz factors. Simulation B15-r10 shows a narrower outflow structure and relatively disordered magnetic field geometry compared to B15-r20, though notably contains similar Lorentz factors. Simulation B5-15-r10, despite forming jets, displays lower Lorentz factors when compared to B15-r20.
The maximum Lorentz factor of B15-r20 is 3.94 whereas for B5-15-r10 it is 2.32. This is likely caused by the jet's radial velocities decreasing over time, as also implied by panels d and e of Fig. \ref{fig:Vr-histograms-mdot}.

\begin{figure*}
\centering
	\begin{minipage}{0.30\textwidth}
	  \includegraphics[width=\textwidth]{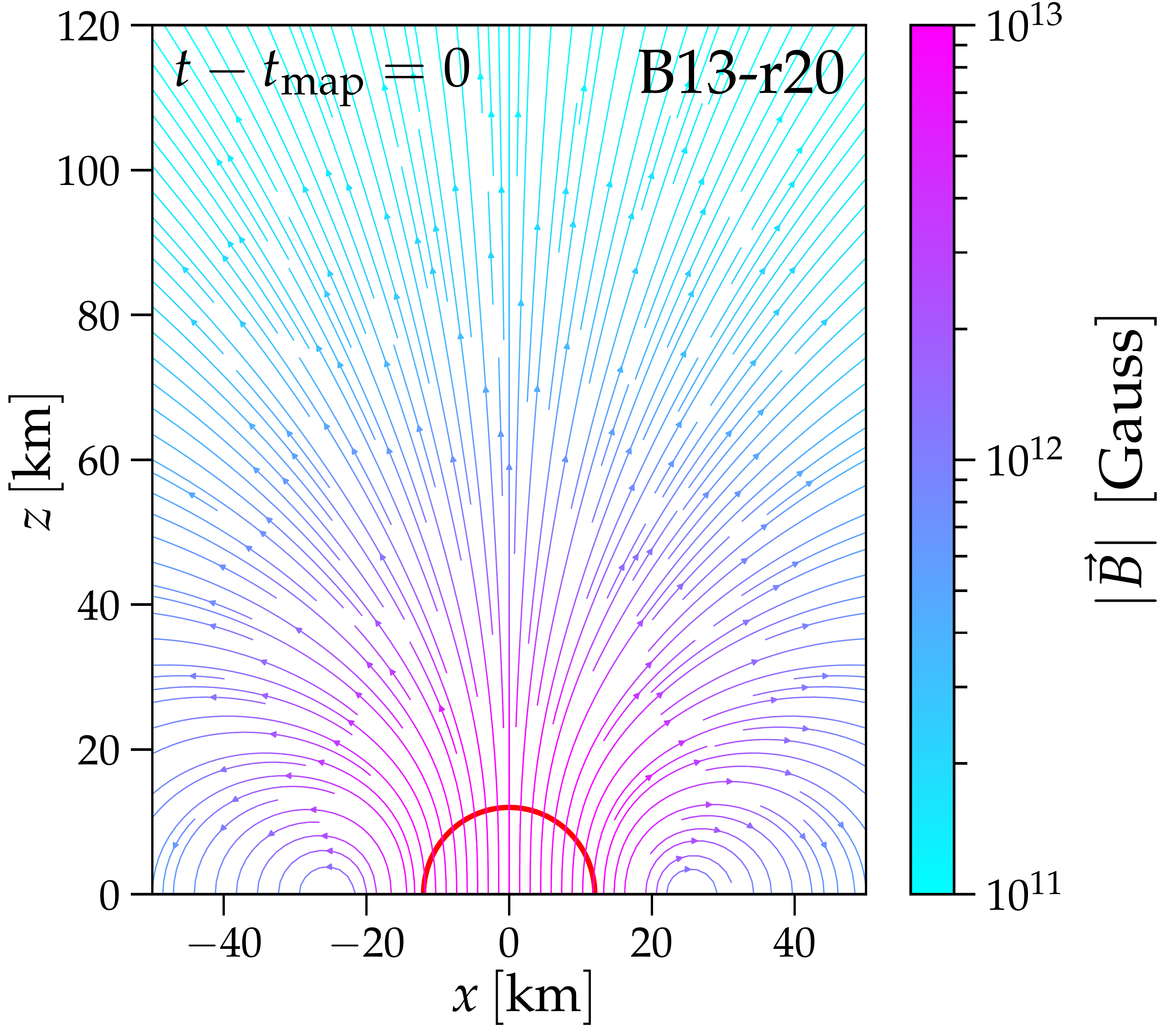}
	\end{minipage}
  \hfill
  \begin{minipage}{0.3\textwidth}
    \includegraphics[width=\textwidth]{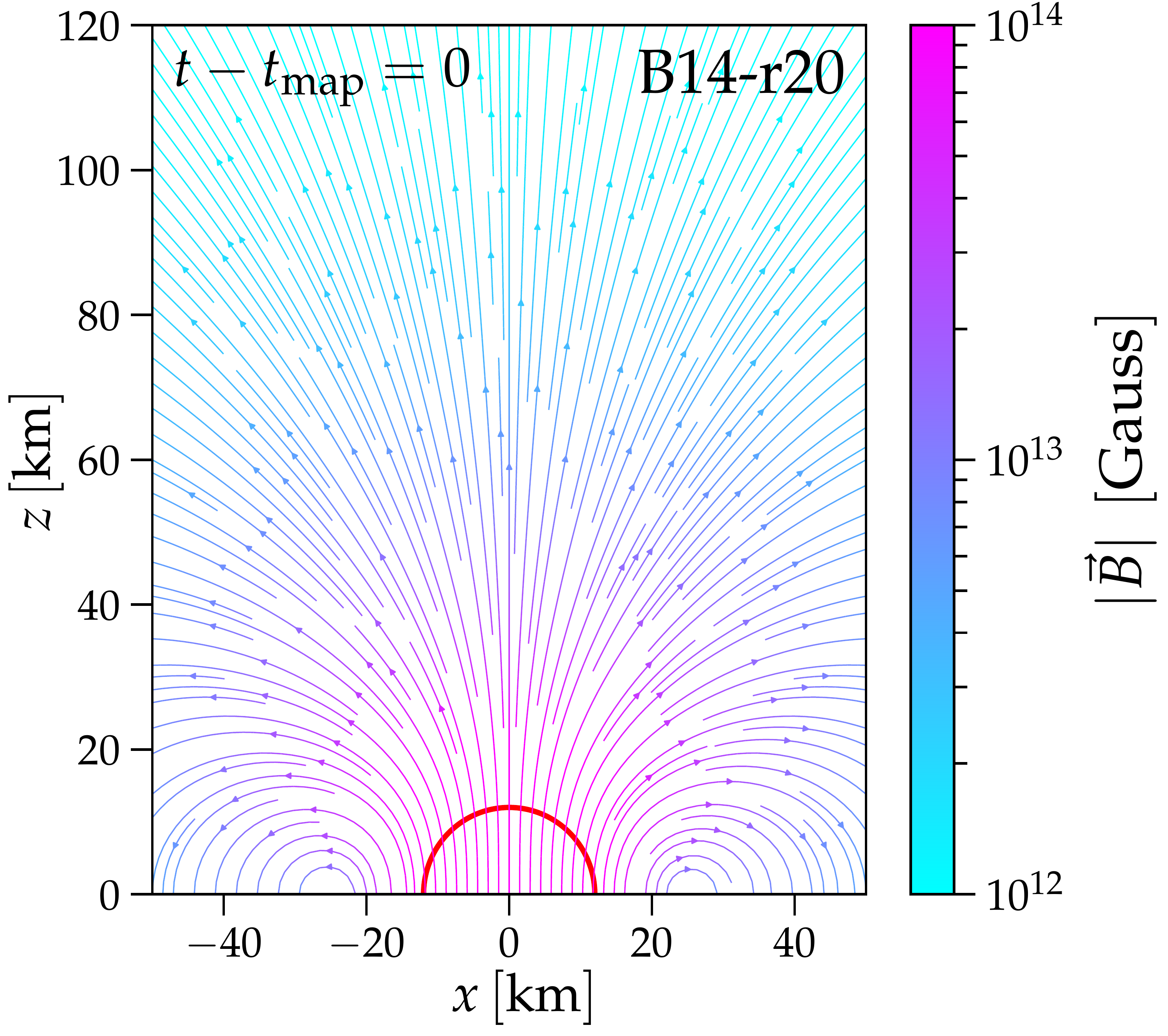}
  \end{minipage}
  \hfill
  \begin{minipage}{0.3\textwidth}
    \includegraphics[width=\textwidth]{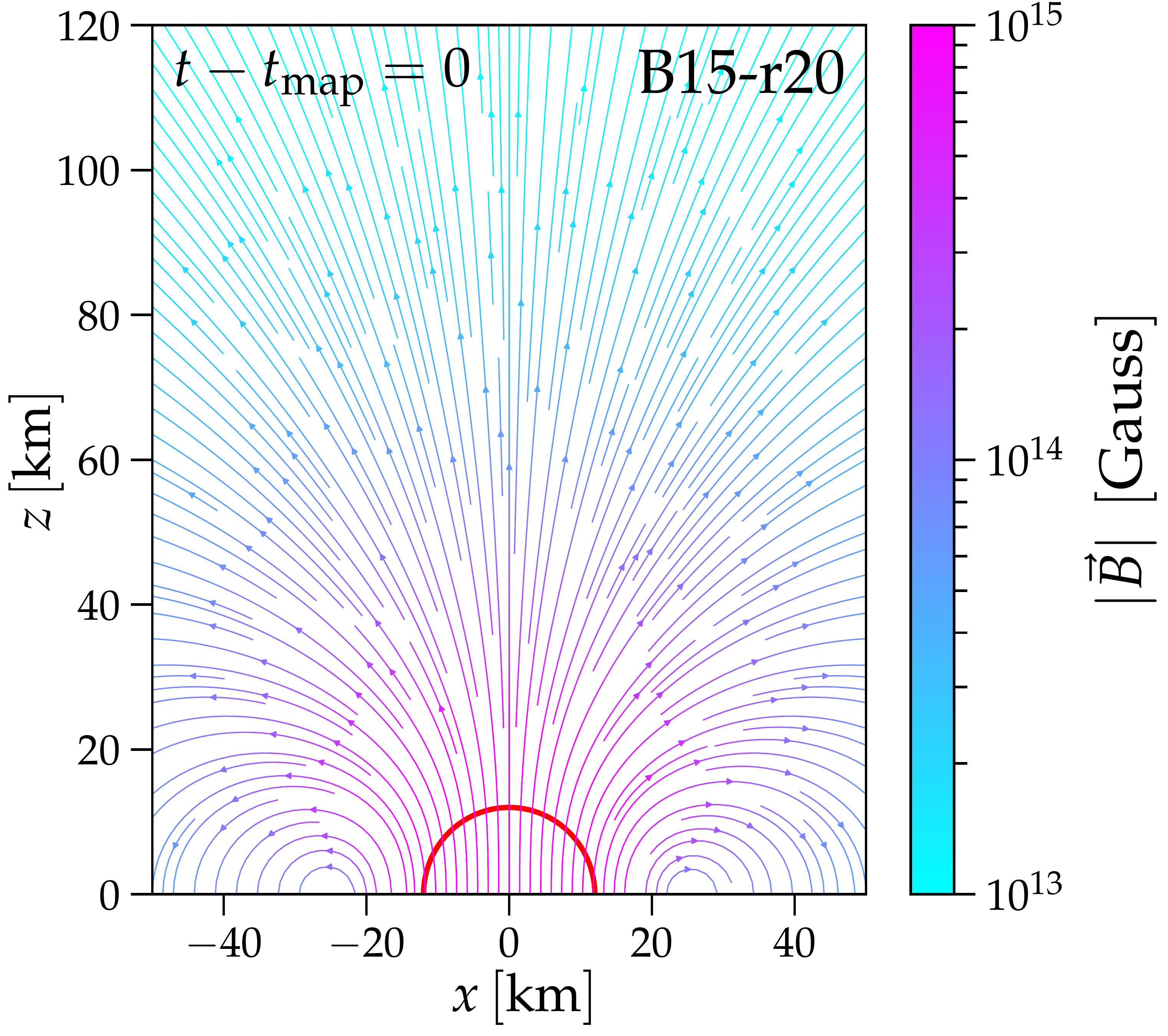}
  \end{minipage} \\
	\begin{minipage}{0.30\textwidth}
	  \includegraphics[width=\textwidth]{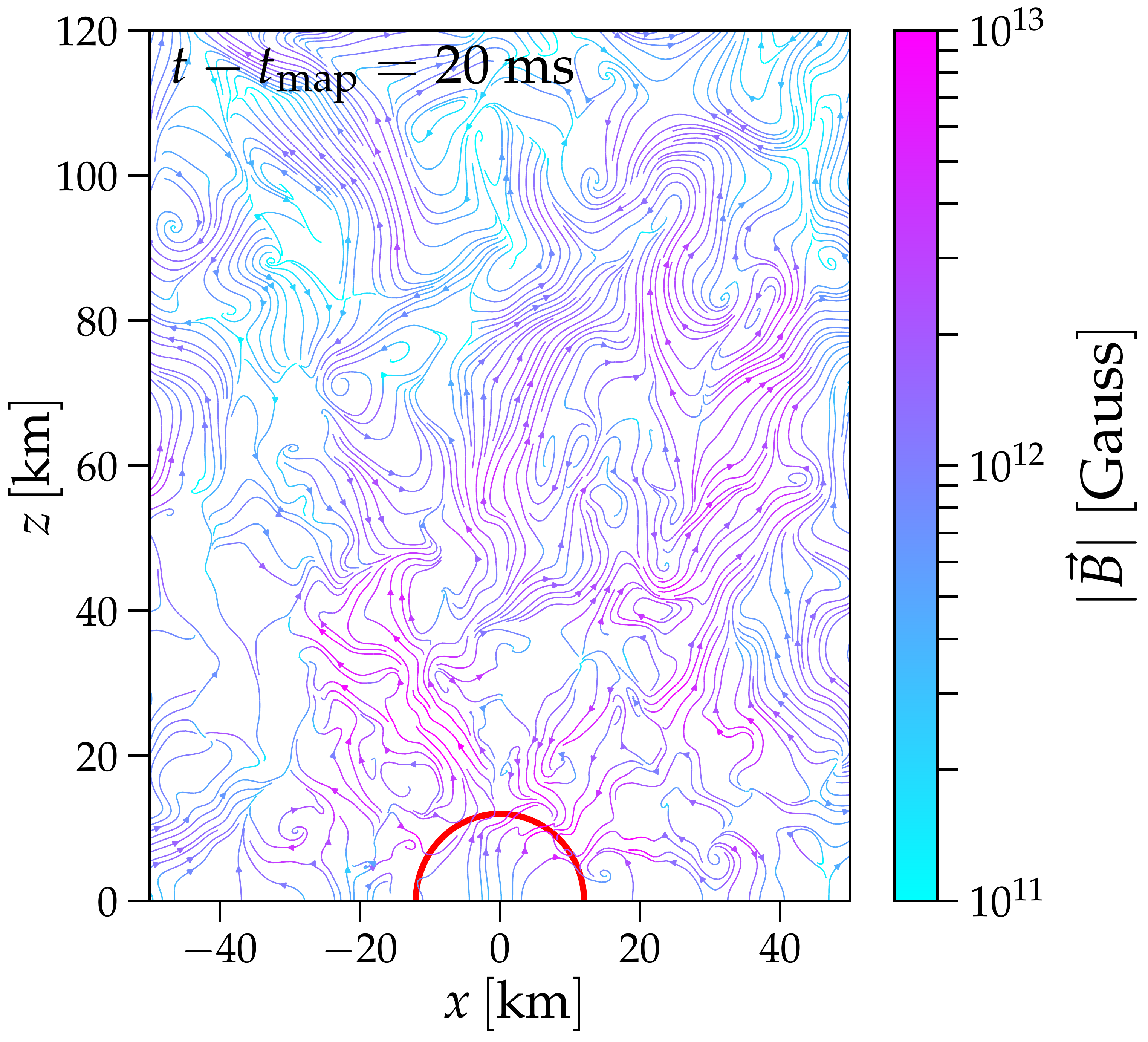}
	\end{minipage}
  \hfill
  \begin{minipage}{0.3\textwidth}
    \includegraphics[width=\textwidth]{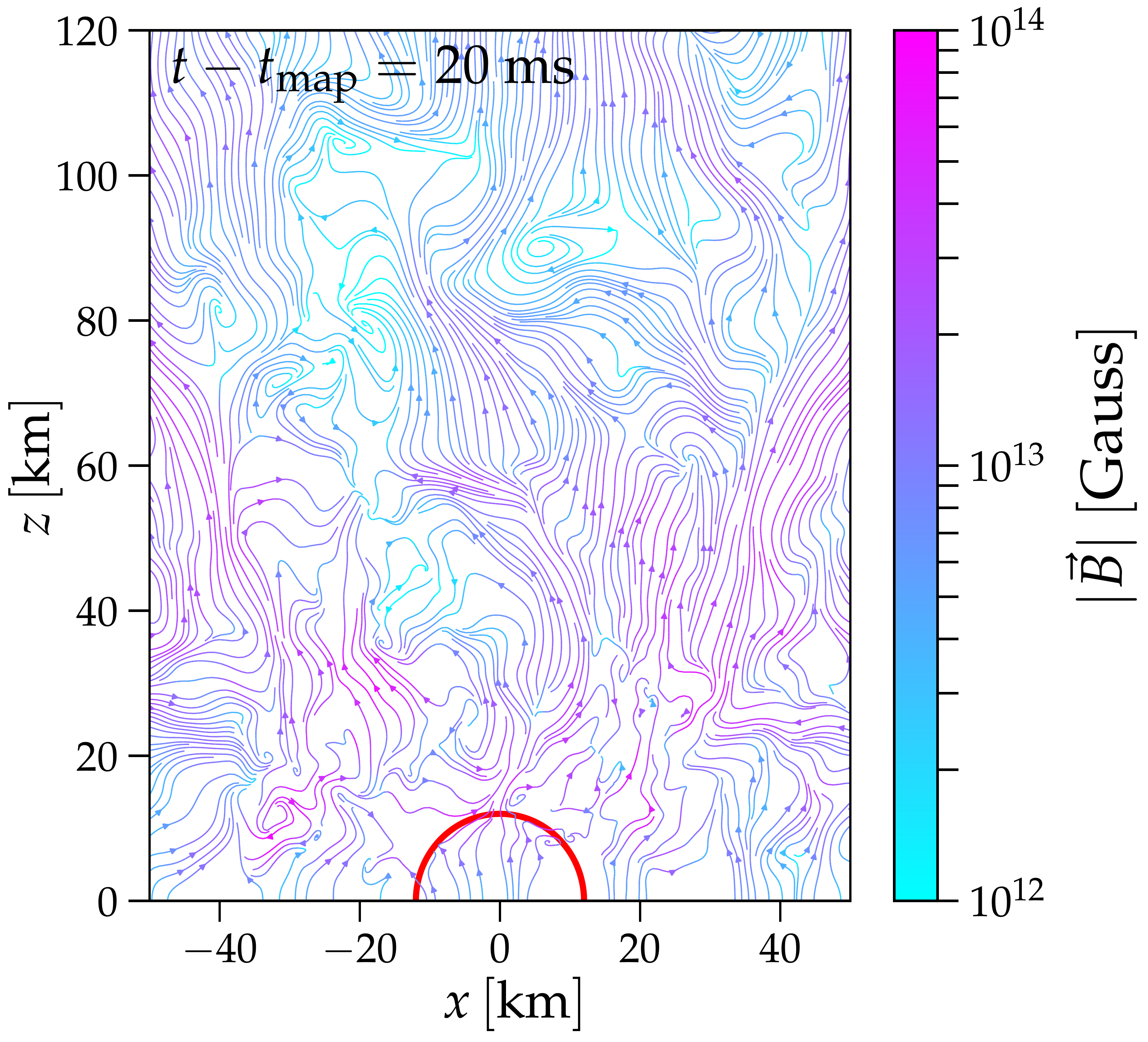}
  \end{minipage}
  \hfill
  \begin{minipage}{0.3\textwidth}
    \includegraphics[width=\textwidth]{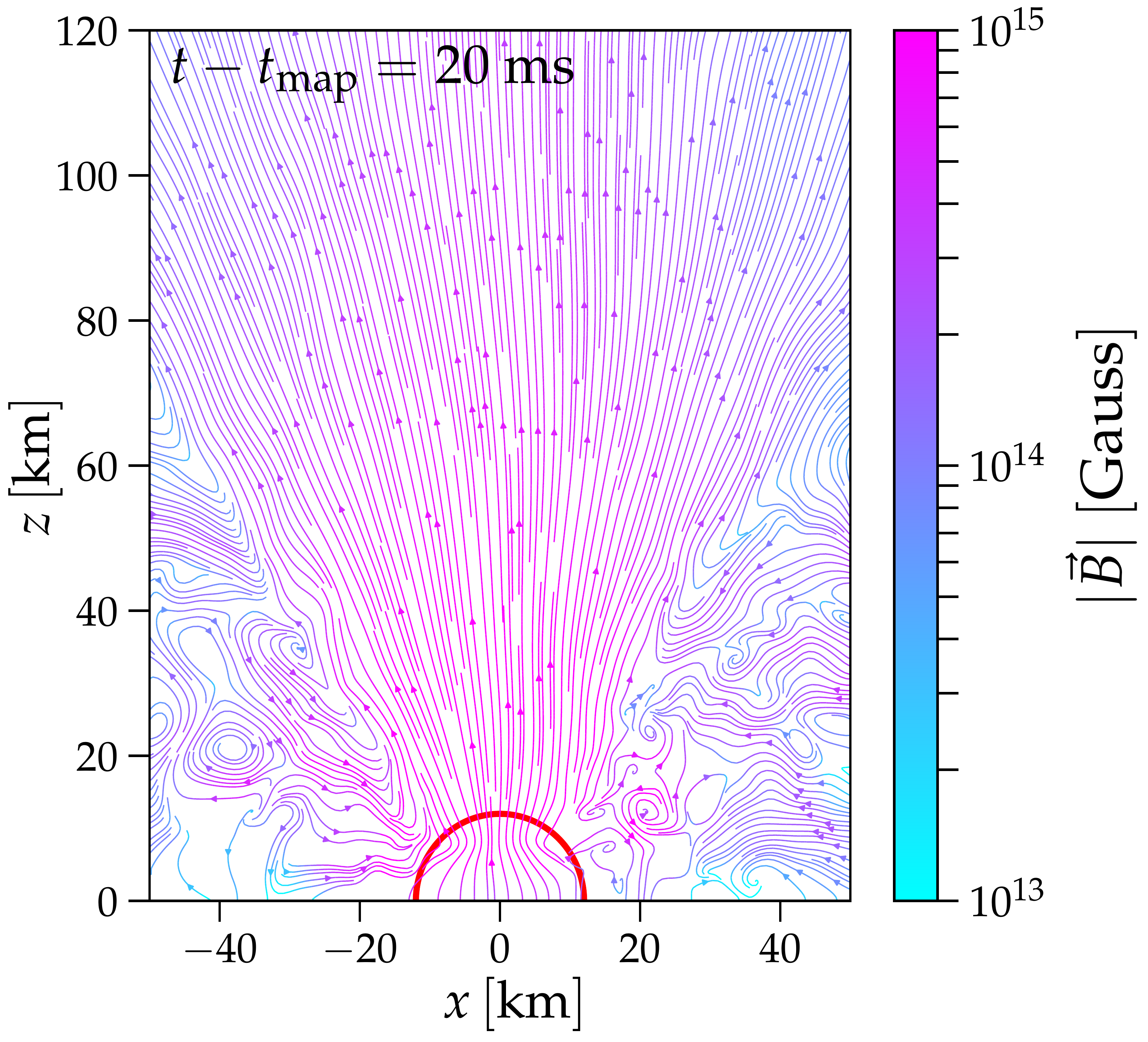}
  \end{minipage}
	\caption{Streamplots of the magnetic field in the meridional ($xz$) plane (where $z$ is the vertical axis) for simulations B13-r20, B14-r20 and B15-r20 at $t - t_{\rm map} = 0$ and 20 ms. For $t - t_{\rm map} = 0$, we compute the magnetic field analytically using the vector potential $A$ in equation \eqref{eq:initialBfield} with varying $B_0$ for each of the displayed simulations. For $t - t_{\rm map} = 20$ ms, we extract the magnetic field from the GRMHD simulations. Note the different limits used for the colorbars.}
	\label{fig:2d-Bvec-B13-14-15} 
\vspace{0.5cm} 
\end{figure*}

\begin{figure*}
\centering
	\begin{minipage}{0.24\textwidth}
	  \includegraphics[width=\textwidth]{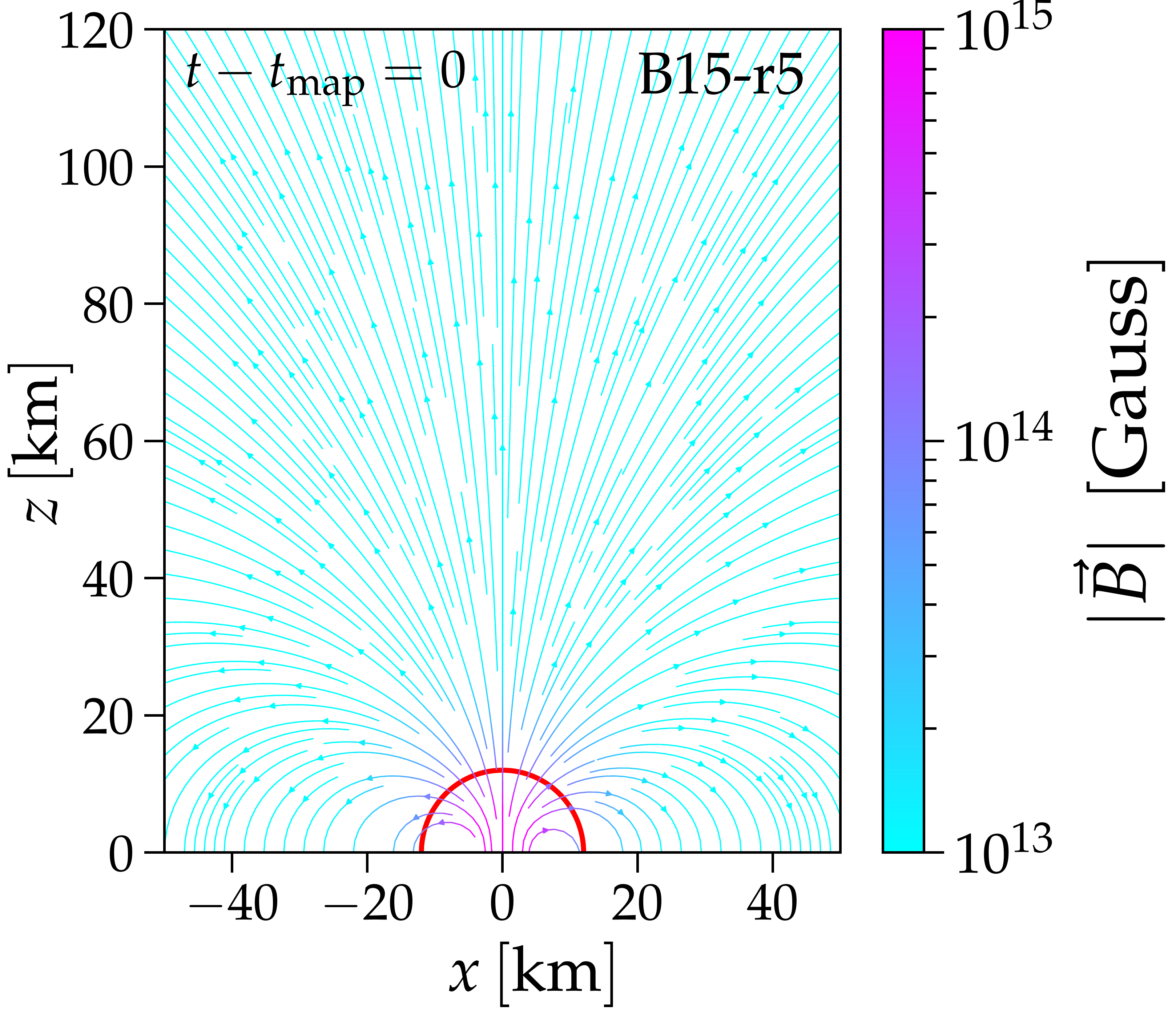}
	\end{minipage}
	\begin{minipage}{0.24\textwidth}
	  \includegraphics[width=\textwidth]{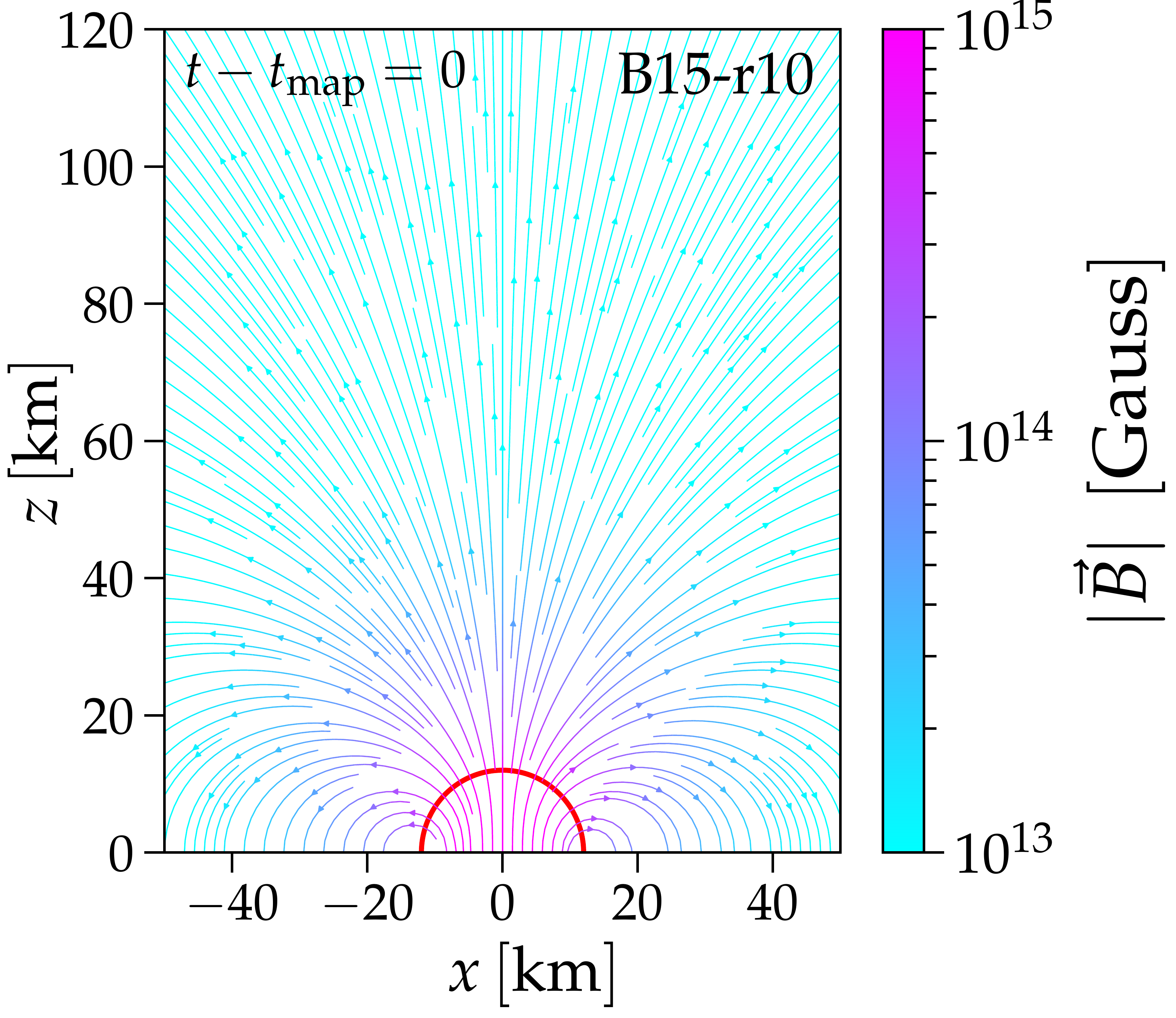} 
	\end{minipage}
	\begin{minipage}{0.24\textwidth}
	  \includegraphics[width=\textwidth]{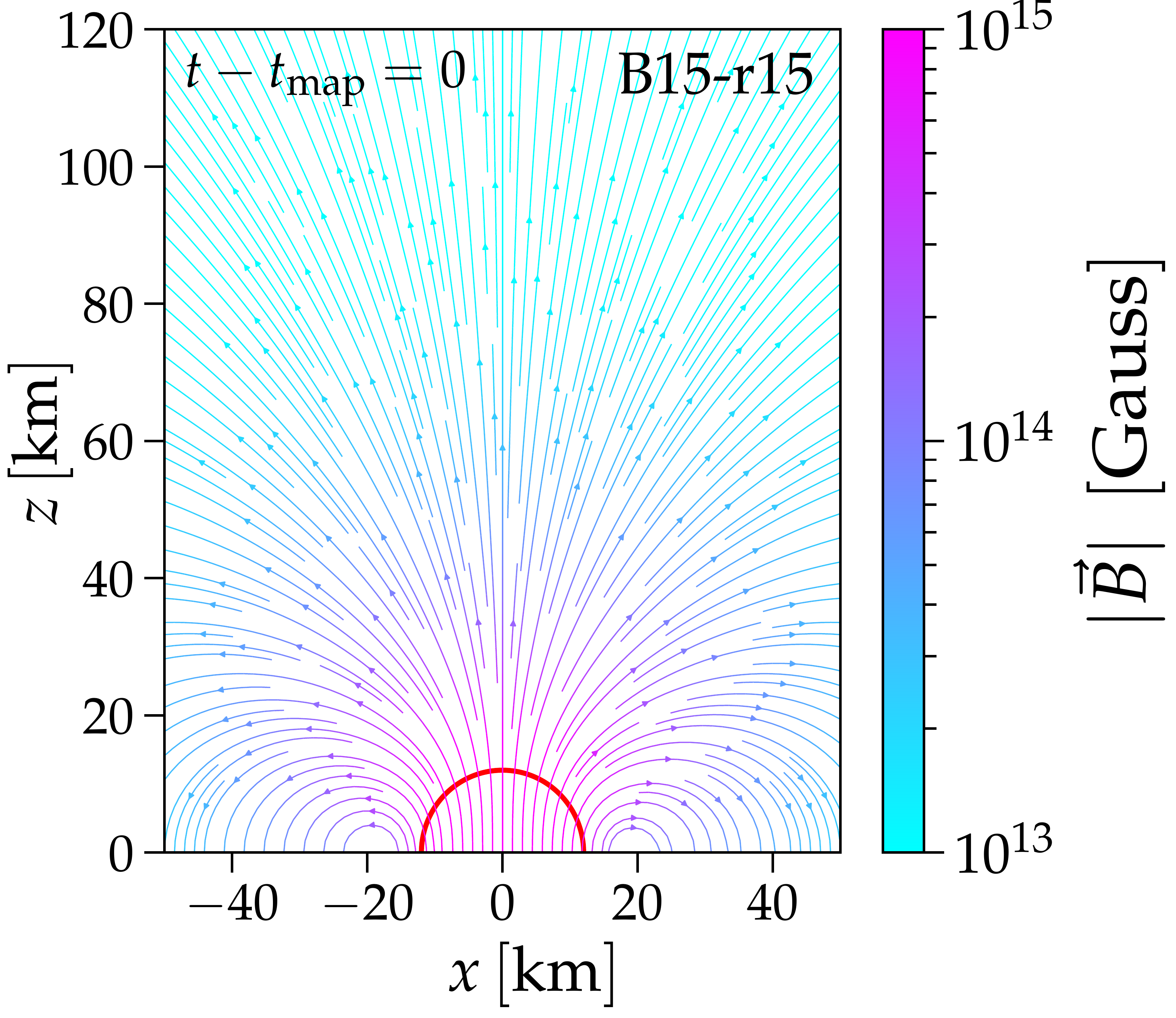} 
	\end{minipage}
	\begin{minipage}{0.24\textwidth}
	  \includegraphics[width=\textwidth]{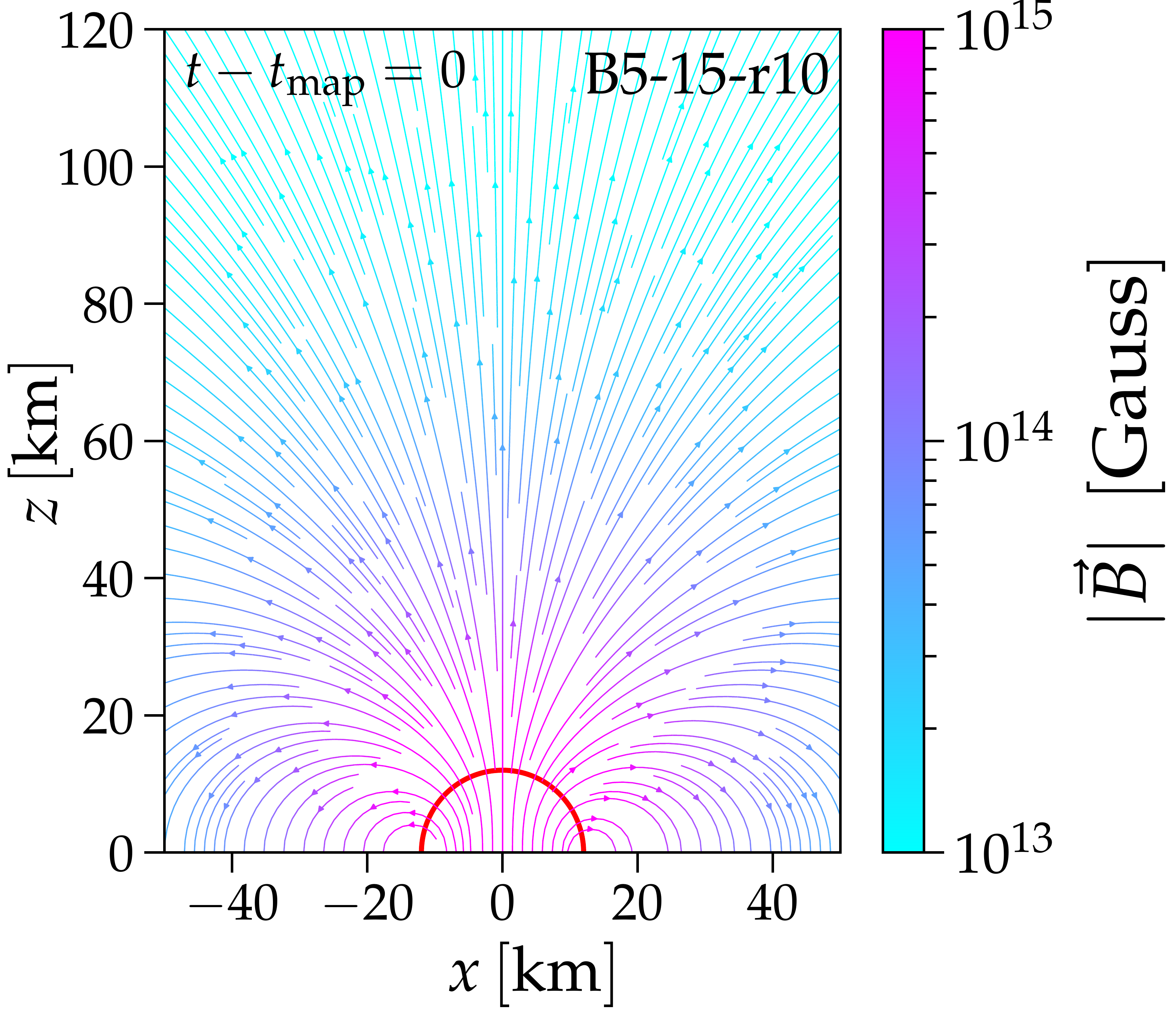}
	\end{minipage}\\
	\begin{minipage}{0.24\textwidth}
	  \includegraphics[width=\textwidth]{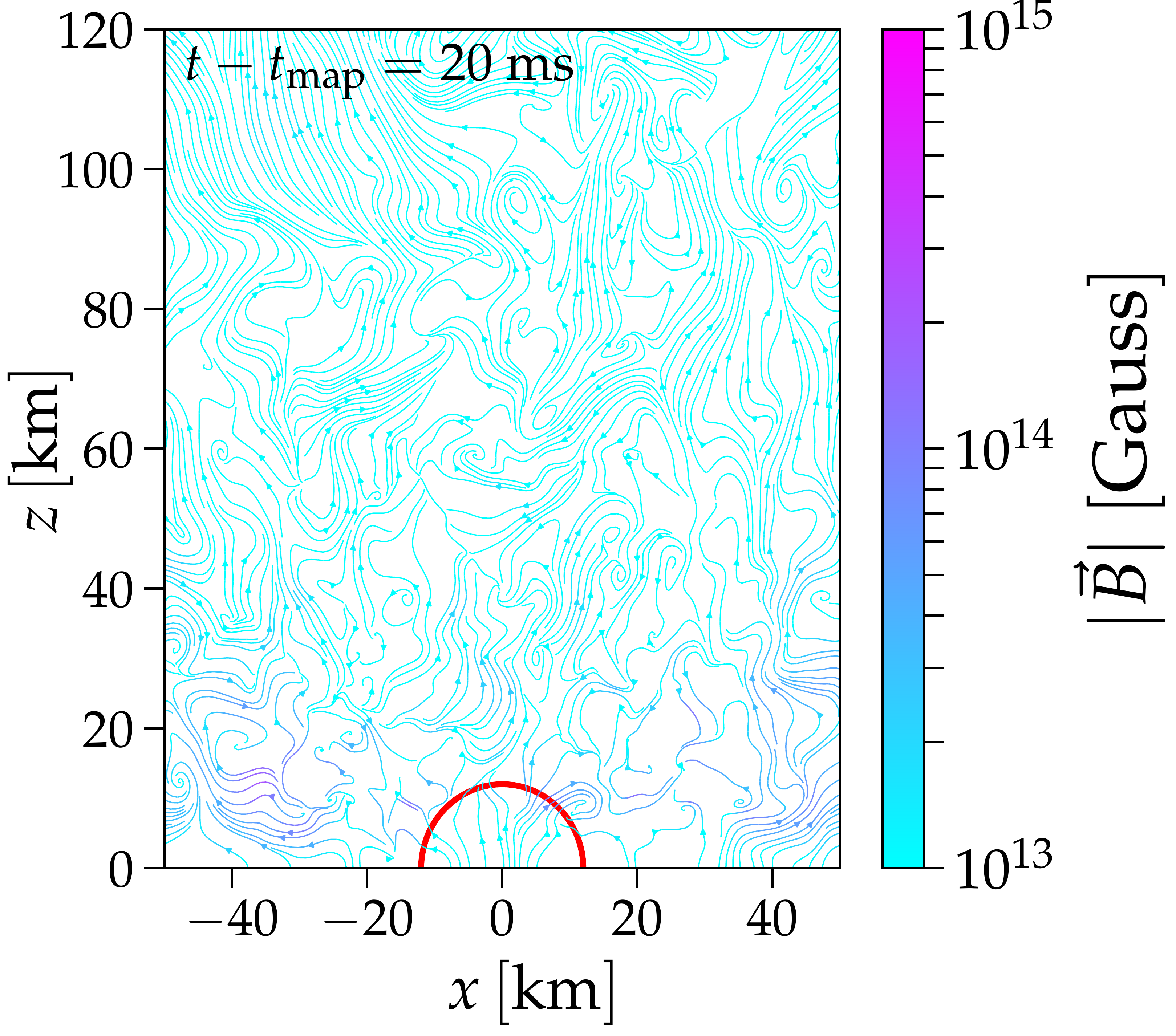}
	\end{minipage}
	\begin{minipage}{0.24\textwidth}
	  \includegraphics[width=\textwidth]{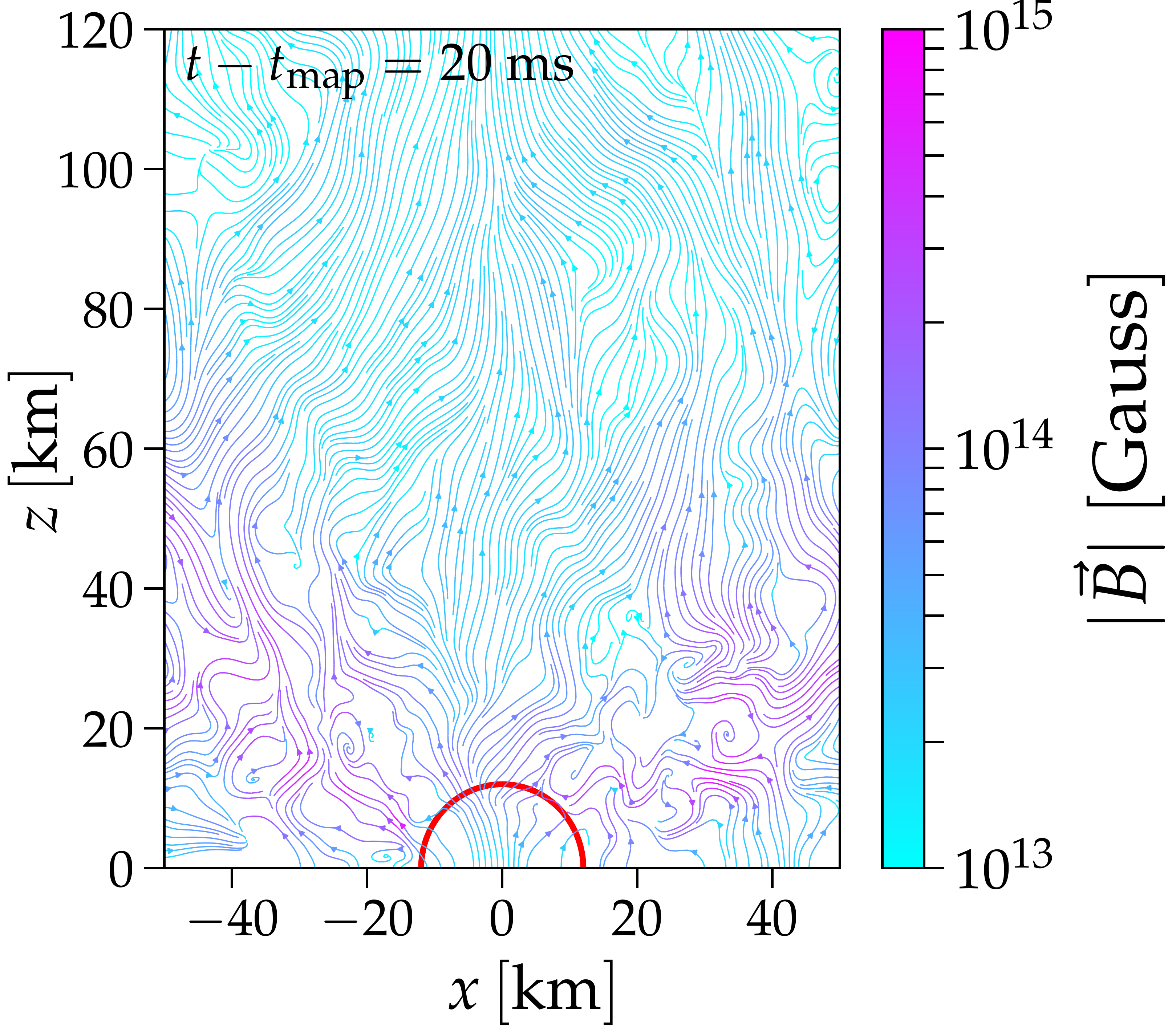} 
	\end{minipage}
	\begin{minipage}{0.24\textwidth}
	  \includegraphics[width=\textwidth]{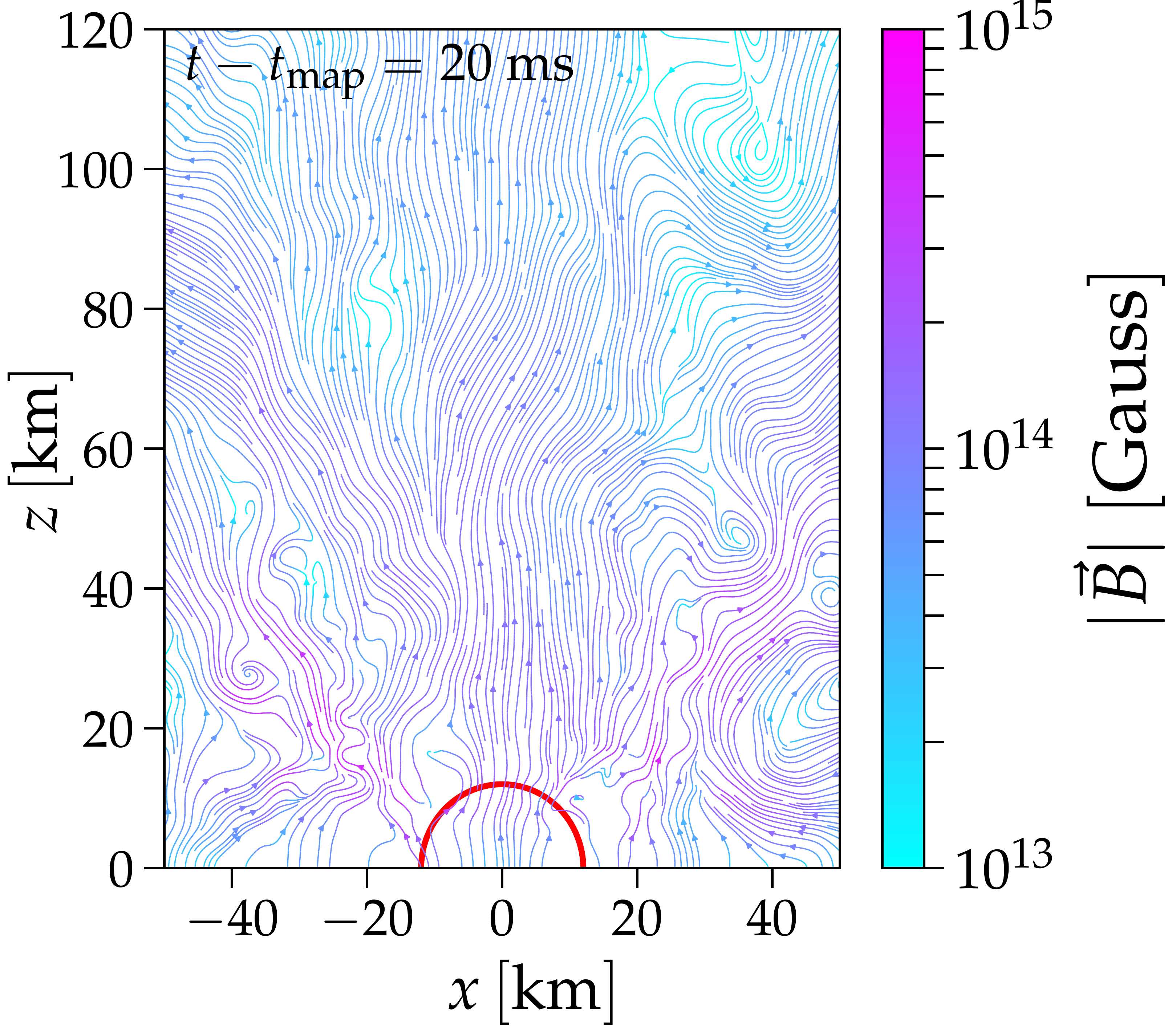} 
	\end{minipage}
	\begin{minipage}{0.24\textwidth}
	  \includegraphics[width=\textwidth]{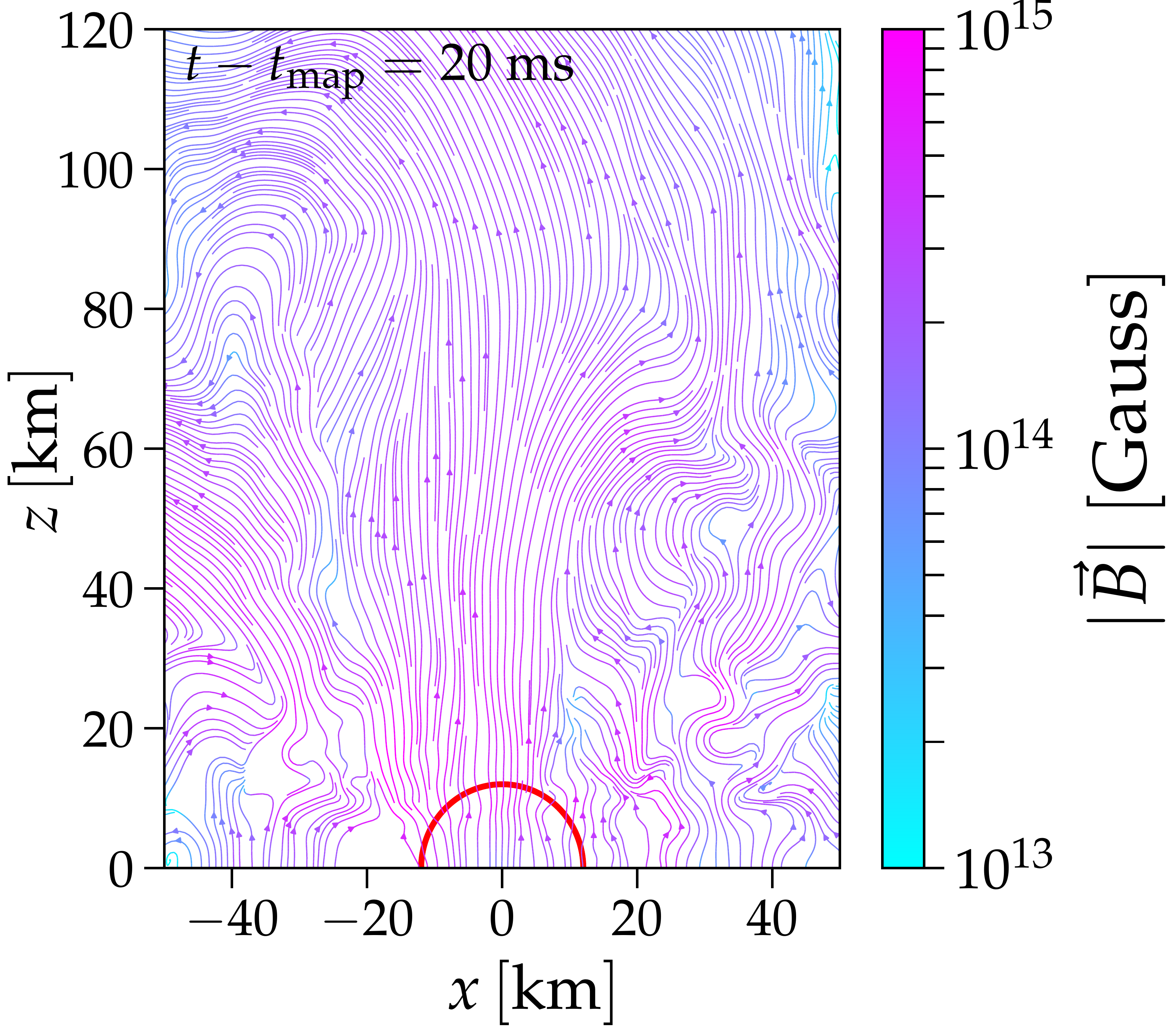}
	\end{minipage}
	\caption{Streamplots of the magnetic field in the meridional ($xz$) plane (where $z$ is the vertical axis) for simulations B15-r5, B15-r10, B15-r15 and B5-15-r10 at $t - t_{\rm map} = 0$ and 20 ms. For $t - t_{\rm map} = 0$, we compute the magnetic field analytically using the vector potential $A$ in equation \eqref{eq:initialBfield} with varying $r_{\rm falloff}$ (and $B_0$ in the case of B5-15-r10) for each of the displayed simulations. For $t - t_{\rm map} = 20$ ms, we extract the magnetic field from the GRMHD simulations.}
	\label{fig:2d-Bvec-falloff} 
\vspace{0.5cm} 
\end{figure*}

\subsection{Evolution of the magnetic field}
\label{sec:magnetic-properties}

In Fig. \ref{fig:2d-Bvec-B13-14-15}, we show streamplots in the meridional ($xz$) plane of the magnetic field (that is, integrating the $\{B_x, B_z\}$ components) for simulations B13-r20, B14-r20 and B15-r20 at $t - t_{\rm map} = 0$ and 20 ms. We adopt three different values for the magnetic field strength $|\vec{B}|$ to highlight their normative features for each simulation. The $t - t_{\rm map} = 0$ magnetic vector field represents the initial ordered magnetic field, which we compute from the vector potential $A$ in equation \eqref{eq:initialBfield} with varying $B_0$ and $r_{\rm falloff} = 20$ for each of the simulations. For $t - t_{\rm map} = 20$ ms, we compute the figures using simulation data, specifically from magnetic variables in the GRMHD evolution of the HMNS system. We infer the relation between the magnetic field parameters and its final configuration by comparing the magnetic field structure at early and late times. This is especially apparent for simulations B13-r20 and B14-r20, which show extreme changes in the magnetic field morphology between $t - t_{\rm map} = 0$ and 20 ms due to the field's adaptation to the underlying magnetohydrodynamical flow of the remnant system, thereby rapidly losing their large-scale structure. For simulation B15-r20, the field appears to be collimated in the polar region due to the development of large toroidal field components, seen in Fig.~\ref{fig:volume-renderings}.

In Fig. \ref{fig:2d-Bvec-falloff}, similarly, we show streamplots in the meridional ($xz$) plane of the magnetic field for simulations B15-r5, B15-r10, B15-r15 and B5-15-r10. The $t - t_{\rm map} = 0$ magnetic vector fields display the initial magnetic field computed from the vector potential $A$ in equation \eqref{eq:initialBfield} with varying $r_{\rm falloff}$ (and $B_0$ for B5-15-r10) for each of the simulations. All simulations, as before, adjust rapidly to the underlying magnetohydrodynamical flow, while showing different magnetic field morphologies and strengths throughout the displayed planes. For simulation B15-r5 at $t - t_{\rm map} = 20$ ms, the magnetic field is dominated by relatively low-$|\vec{B}|$ values and disordered field configurations. Simulations B15-r10 and B15-r15, by contrast, display larger magnetic field strengths and higher degrees of order in their field structures, albeit also exhibiting disordered and/or low-$|\vec{B}|$ regions. Simulation B5-15-r10 exhibits the most ordered field in combination with high-$|\vec{B}|$ regions, although notably showing a considerably different field morphology compared to B15-r20 in Fig. \ref{fig:2d-Bvec-B13-14-15}.

\subsection{Nucleosynthesis and kilonovae}
\label{sec:nucleo-and-kilonovae}

\begin{figure*}
\centering
	\begin{minipage}{0.47\textwidth}
	\includegraphics[width=\textwidth]{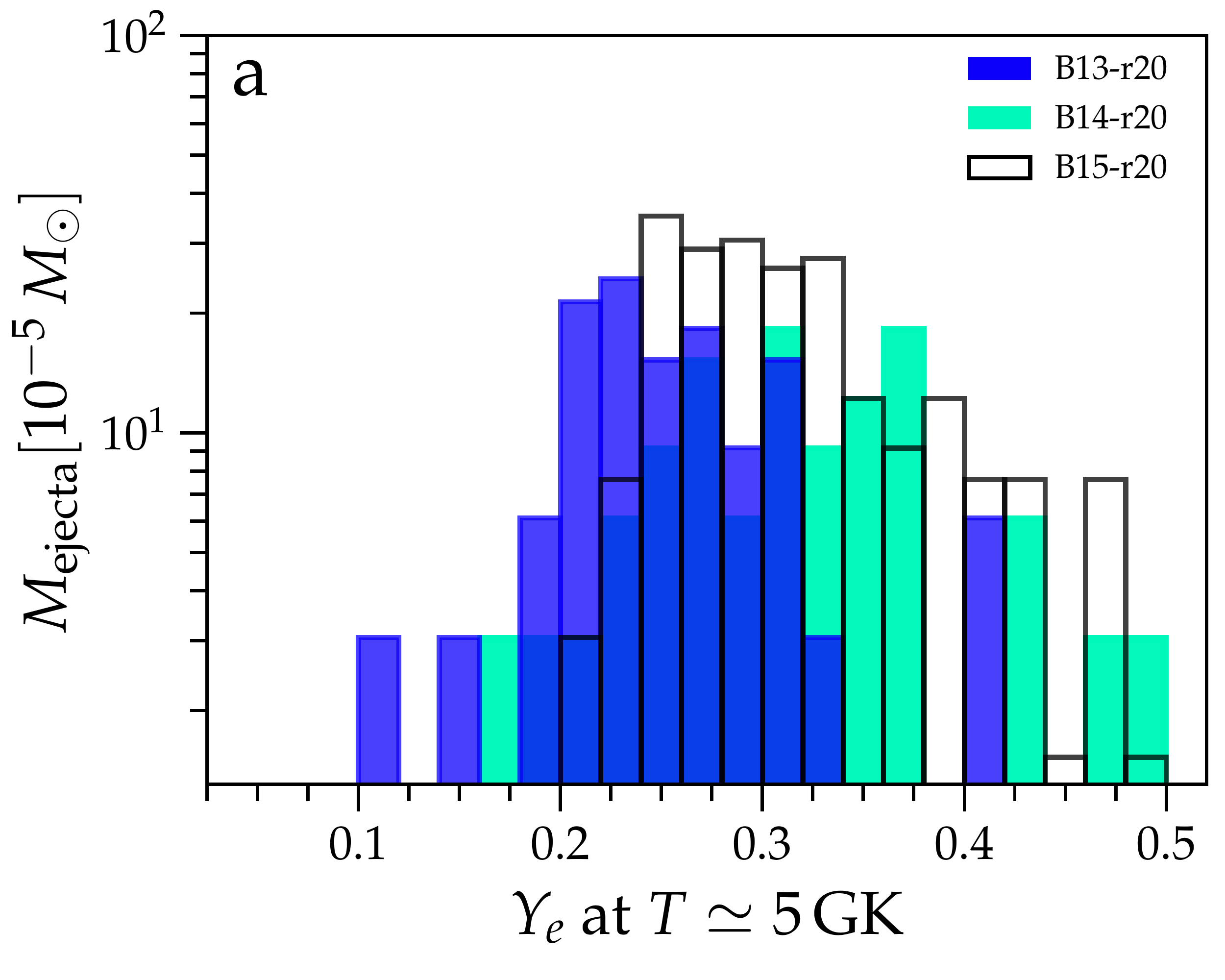}
	\end{minipage}
	\begin{minipage}{0.47\textwidth}
	\hspace{-0.27cm}
	\includegraphics[width=\textwidth]{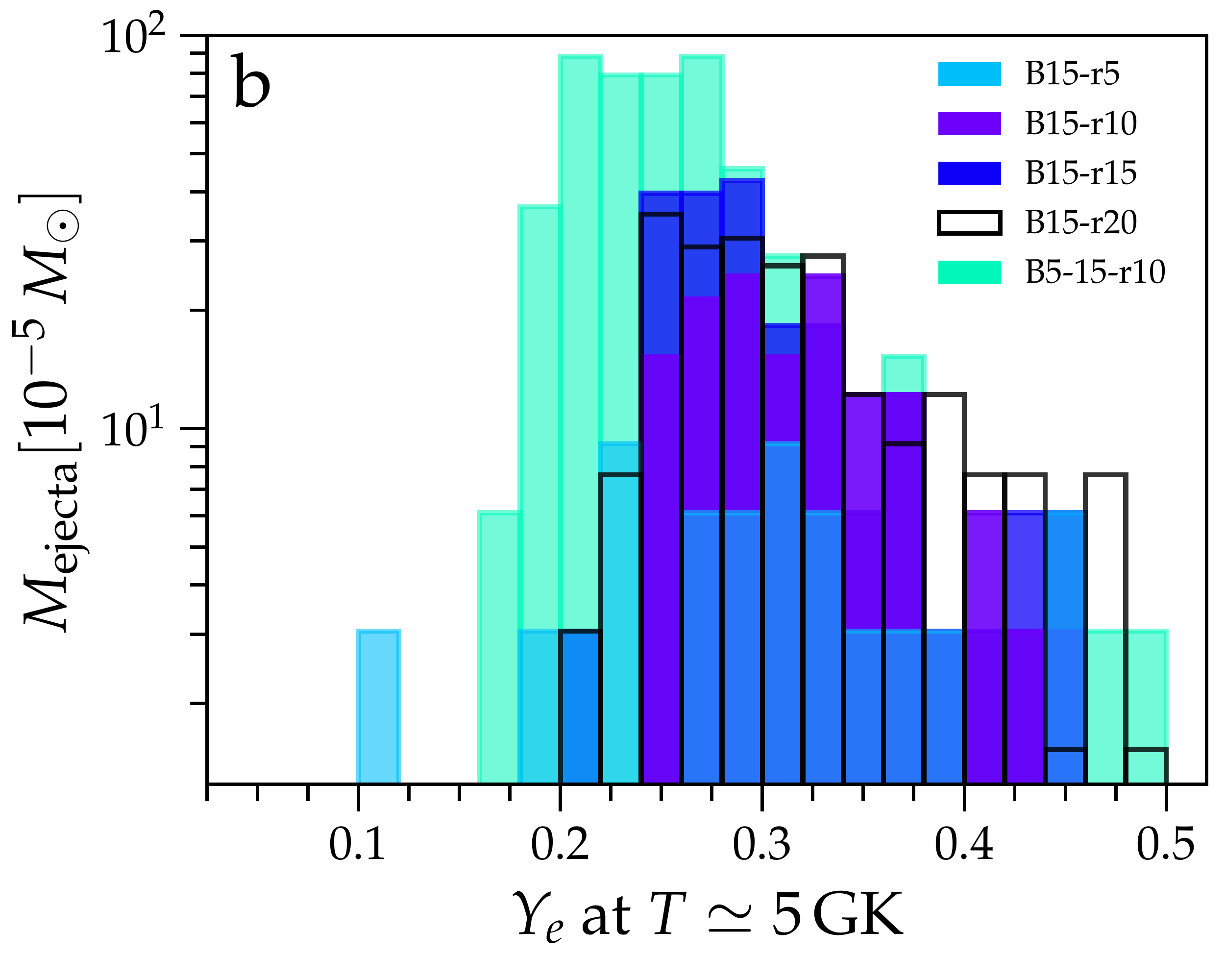} 
	\end{minipage}
	\caption{Panel a: $Y_e$ histograms of all tracer particles for B13-r20 (blue), B14-r20 (green) and B15-r20 (black), when the temperature of the particles is last above 5 GK. We compute the $Y_e$ distributions using \texttt{SkyNet}. Simulation B15-r20 is shown here and in panel b in black, as it is nearly identical to the lowest-resolution simulation of \citet{2020ApJ...901L..37M}. Panel b: $Y_e$ histograms of all tracer particles for B15-r5 (cyan), B15-r10 (purple), B15-r15 (blue), B15-r20 (black) and B5-15-r10 (green) when the temperature of the particles is last above 5 GK, which we again compute using \texttt{SkyNet}.}
	\label{fig:Ye5GK} 
\vspace{0.5cm} 
\end{figure*}

In panel a of Fig. \ref{fig:Ye5GK}, we show electron fraction histograms of all tracer particles for simulations B13-r20, B14-r20 and B15-r20, when the temperature of the particles is last above 5 GK. As this is approximately the temperature at which $r$-process nucleosynthesis starts, the electron fractions at this temperature are the relevant quantities for setting the $r$-process yields. As mentioned, the approximate neutrino scheme of the simulations causes uncertainties in our nucleosynthesis predictions, where the $r$-process production of heavy elements may be reduced by up to a factor of $\sim$10 \citep[when comparing the most extreme cases;][]{2021arXiv211200772C}. We compute the $Y_e$ distributions using \texttt{SkyNet} \citep{2017ApJS..233...18L}. All simulations exhibit wide distributions in $Y_e$, where especially B13-r20 contains more low-$Y_e$ material while also showing some ejecta in the $0.1 < Y_e < 0.16$ range. Simulation B14-r20 contains significant $Y_e > 0.4$ ejecta, while also displaying a larger average $Y_e$ compared to B13-r20. For B15-r20, a large amount of $0.24 < Y_e < 0.34$ material is ejected, while also showing significant high-$Y_e$ material. These results seem to tentatively imply that when increasing $B_0$, the $Y_e$ of the ejecta generally shifts to larger values. 

In panel b of Fig. \ref{fig:Ye5GK}, we show $Y_e$ distributions of all tracer particles when their temperature is last above 5 GK for simulations B15-r5, B15-r10, B15-r15, B15-r20 and B5-15-r10. Simulation B15-r5 mostly contains ejecta with $Y_e \sim 0.3$, albeit also showing both low- and high-$Y_e$ material including an extremely low electron fraction at $0.1 < Y_e < 0.12$. Simulations B15-r10 and B15-r15 mostly show ejecta around $Y_e \sim 0.3$, although displaying significantly shallower distributions compared to the other simulations. Simulation B15-r20 contains similar ejecta masses compared to B15-r10 and B15-r15 around $Y_e \sim 0.3$, although showing a considerably wider distribution. For B5-15-r10, a lower peak around $Y_e \sim 0.24$ and significant low-$Y_e$ ejecta of $Y_e < 0.2$ is inferred. Although $r_{\mathrm{falloff}}$ comes in with a cubic power in Eq.~\eqref{eq:initialBfield}, due to the astrophysically-relevant small parameter range used, it is harder to discern a clear trend between $r_{\mathrm{falloff}}$ and $Y_e$. Indeed, some of the histograms have broadly similar features, which is to be expected given that the changes introduced through $r_{\mathrm{falloff}}$ are slightly more subtle. The differences in the $Y_e$ distribution could arise due to the variation of the falloff parameter and/or differences in the flow structure that individual tracer particles advect along.

\begin{figure*}
\centering
	\begin{minipage}{0.47\textwidth}
	\includegraphics[width=\textwidth]{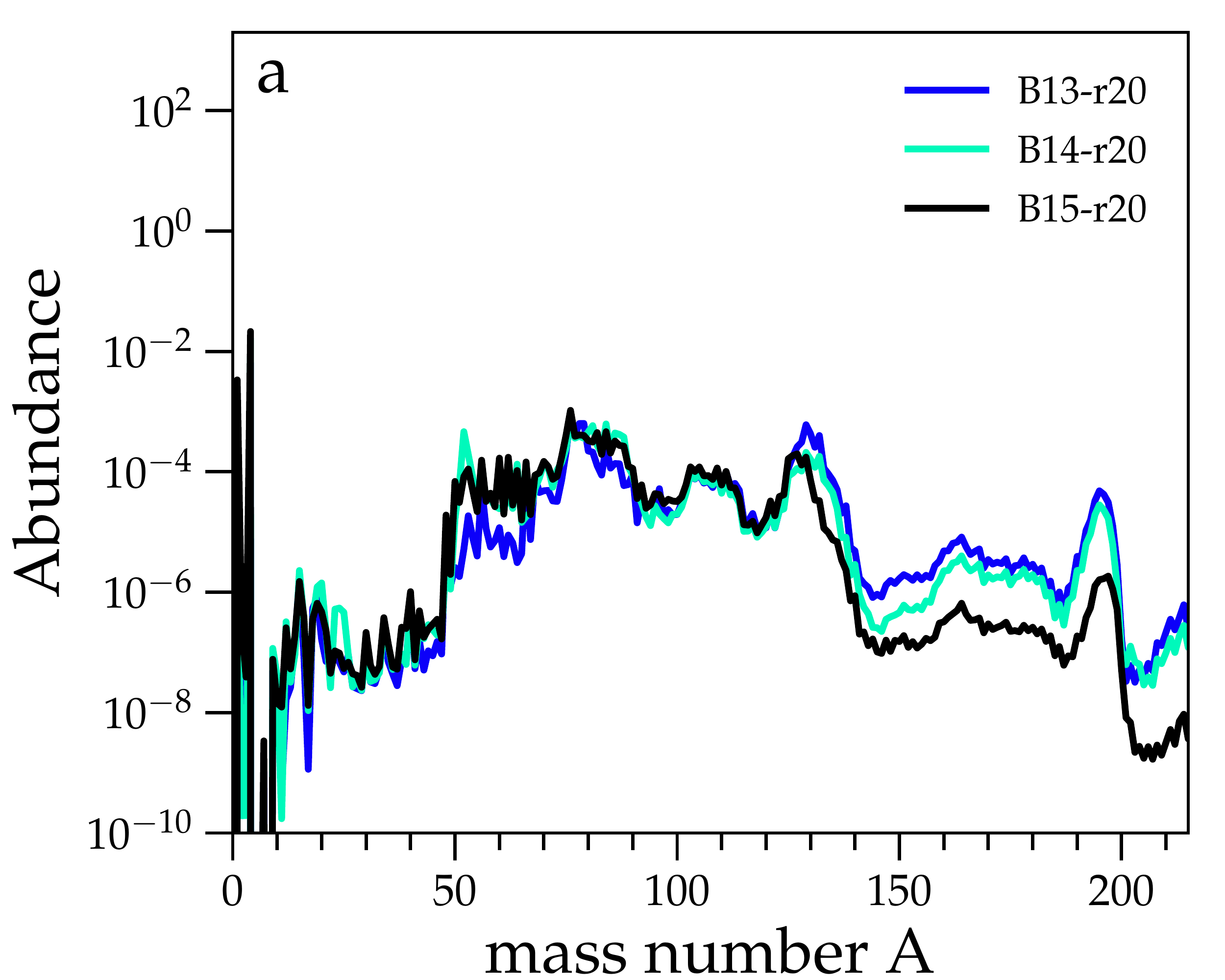}
	\end{minipage}
	\begin{minipage}{0.47\textwidth}
	\hspace{-0.27cm}
	\includegraphics[width=\textwidth]{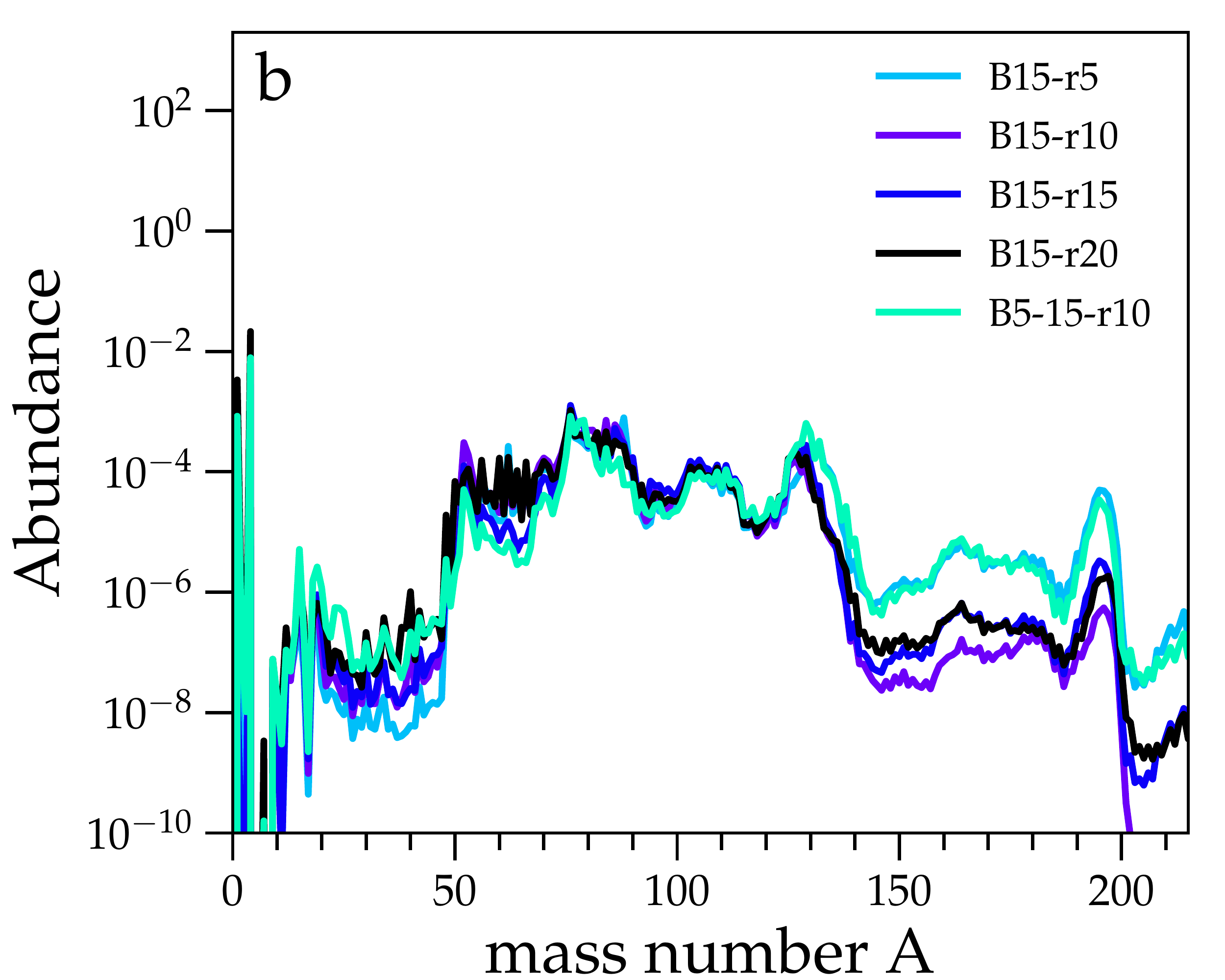} 
	\end{minipage}
	\caption{Panel a: Fractional abundances vs mass number for simulations B13-r20 (blue), B14-r20 (green) and B15-r20 (black), which we compute using the recorded neutrino luminosities of tracer particles. Simulation B15-r20 is shown here and in panel b in black, as it is nearly identical to the lowest-resolution simulation of \citet{2020ApJ...901L..37M}. Panel b: Fractional abundances vs mass number for simulations B15-r5 (cyan), B15-r10 (purple), B15-r15 (blue), B15-r20 (black) and B5-15-r10 (green), which we again compute using the encountered neutrino luminosities of tracer particles.}
	\label{fig:abundancesall} 
\vspace{0.5cm} 
\end{figure*}

In Panel a of Fig. \ref{fig:abundancesall}, we show the fractional abundances as a function of mass number for simulations B13-r20, B14-r20 and B15-r20. We compute these abundances using the neutrino luminosity recorded by tracer particles for each simulation. As mentioned, the $Y_e$ distributions (see Fig. \ref{fig:Ye5GK}) for each simulation should coincide with the inferred abundances, where $Y_e \lesssim 0.2$ ejecta causes a strong $r$-process, $0.25 \lesssim Y_e \lesssim 0.4$ results in unsubstantial amounts of heavy nuclei ($A > 140$) production and $Y_e \gtrsim 0.4 - 0.5$ causes a weak $r$-process \citep{2021arXiv211200772C}. It is mainly interesting to investigate the amount of heavy nuclei production, for which B13-r20 shows the largest abundances for the majority of mass numbers. Simulation B14-r20 shows similar abundances in the heavy-nuclei regime, except in the range of $140 < A < 155$. For B15-r20, considerably lower amounts of heavy nuclei are produced for nearly all $A > 140$ regimes. The fractional abundances of these three simulations seem to be in line with the $Y_e$ distributions in Panel a of Fig. \ref{fig:Ye5GK}, as a larger $B_0$ leads to a decrease in heavy element production.

In Panel b of Fig. \ref{fig:abundancesall}, we show the fractional abundances as a function of mass number for simulations B15-r5, B15-r10, B15-r15, B15-r20 and B5-15-r10. We compute the abundances using the neutrino luminosities encountered by tracer particles. Notably, B15-r5 and B5-15-r10 display very similar abundances for $A > 140$, while also producing the largest fractions of heavy elements when compared to the other simulations in this panel. Indeed, the ejected material for simulations B15-r5 is only sampled by a small amount of tracers particles, which give rise to an abundance computation based on relatively low statistics. This may impact the relative abundances tracers are probing. For B15-r20 and B15-r15, similar heavy nuclei production is inferred, albeit not forming significant amounts of $A > 140$ material. Simulation B15-r10 displays even less nuclei with $A > 140$, while its fractional abundance rapidly drops after $A \gtrsim 200$. The abundances and the $Y_e$ distribution are correlated as we expected, however, a definitive trend between the abundances and $r_{\mathrm{falloff}}$ is hard to discern. This, similarly, could come down to the trajectories of tracer particles within each simulations or to the subtle impact of $r_{\mathrm{falloff}}$ on the outflow composition.

\begin{figure*}
\centering
	\begin{minipage}{0.44\textwidth}
	\includegraphics[width=\textwidth]{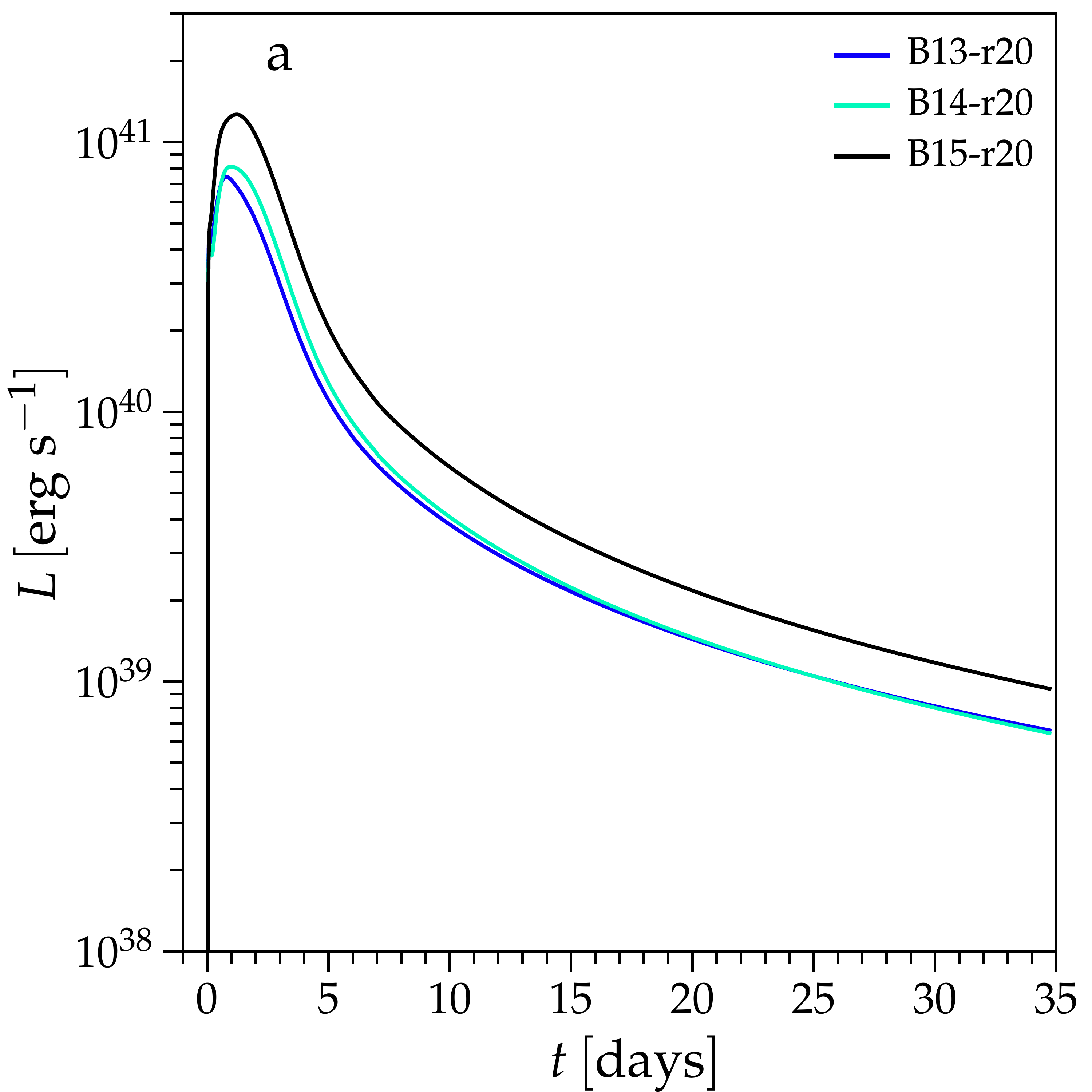}
	\end{minipage}
	\begin{minipage}{0.44\textwidth}
	\hspace{-0.27cm}
	\includegraphics[width=\textwidth]{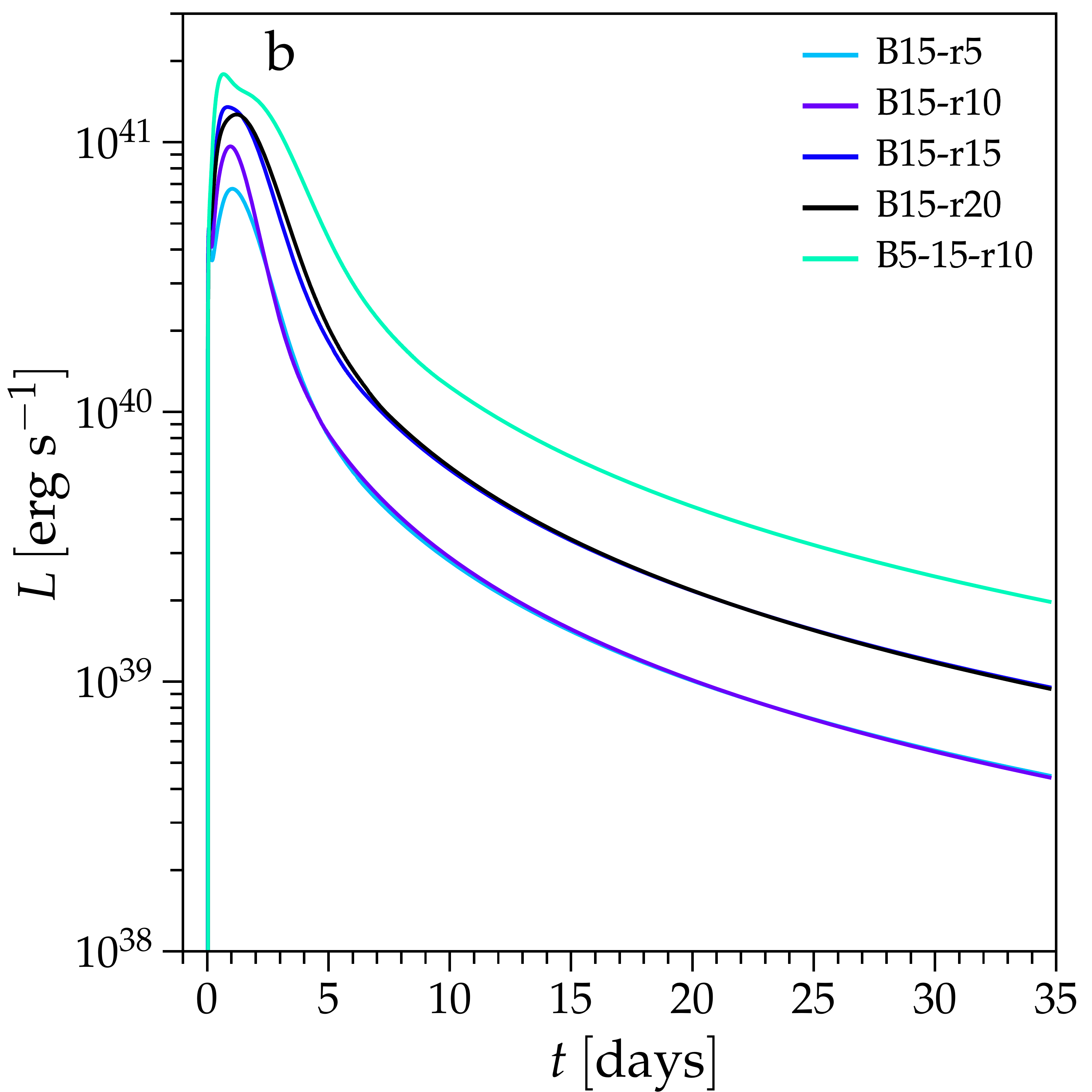} 
	\end{minipage}
	\caption{Panel a: Bolometric luminosity (computed at a distance $r = 100\, M_{\odot}$) as a function of time for simulations B13-r20 (blue), B14-r20 (green) and B15-r20 (black). Simulation B15-r20 is shown here and in panel b in black, as it is nearly identical to the lowest-resolution simulation of \citet{2020ApJ...901L..37M}. Panel b: Bolometric luminosity (computed at a distance $r = 100\, M_{\odot}$) as a function of time for simulations B15-r5 (cyan), B15-r10 (purple), B15-r15 (blue), B15-r20 (black) and B5-15-r10 (green).}
	\label{fig:inferredkilonovae} 
\vspace{0.5cm} 
\end{figure*}

In Fig. \ref{fig:inferredkilonovae}, we show kilonova light curves in terms of the bolometric luminosities $L$ for all simulations, which we compute using outflow properties extracted at a radius of $r = 100 M_{\odot}$. In Panel a, we show the bolometric luminosities for simulations with varying $B_0$. Simulations B13-r20 and B14-r20 exhibit very similar light curves, where the latter shows a slightly brighter peak. Simulation B15-r20 contains significantly larger luminosity values throughout its evolution compared to B13-r20 and B14-r20.

In Panel b of Fig. \ref{fig:inferredkilonovae}, we show the bolometric luminosities obtained at $r = 100 M_{\odot}$, in this case for simulations B15-r5, B15-r10, B15-r15, B15-r20 and B5-15-r10. The brightest kilonova is produced by B5-15-r10, which shows both the largest luminosity peak and consistently larger $L$ compared to the other simulations, including B15-r20. For B15-r15 and B15-r20, very similar kilonova light curves and peak values are obtained. Simulations B15-r5 and B15-r10 also exhibit similar luminosity evolution, although the latter produces a significantly larger peak.

\section{Summary and conclusions}
\label{ch:summary-conclusions}

We have performed seven GRMHD simulations of a HMNS system with varying parameterized magnetic field strengths and configurations, to investigate its effects on the outflow properties, nucleosynthesis yields and kilonova light curves. Our simulations include a neutrino treatment and tabulated, nuclear EOS. 

Simulations B15-r20 and B5-15-r10, which contain the strongest magnetic fields, show the emergence of collimated, mildly-relativistic jets as opposed to magnetized winds only. Jets can emerge in the simulations as a result of the strong magnetic fields in addition to the incorporation of neutrino effects, as this reduces baryon pollution in the polar regions \citep[e.g.,][]{2020ApJ...901L..37M}. The jets are then collimated by hoop stresses from the strong toroidal magnetic field windup along the rotation axis of the remnant. For B5-15-r10 and B15-r20, we find multiple indications for the presence of mildly-relativistic jets. Most notably, these two simulations exhibit larger velocities of unbound material and mass ejecta rates (see Fig. \ref{fig:Vr-histograms-mdot}) compared to the other simulations. Moreover, the earlier collapse times of B5-15-r10 and B15-r20 (by $\sim 1.6$ ms, see Fig. \ref{fig:rhomaxvstime}) indicate that angular momentum is extracted more efficiently from the remnant, pointing towards jetted outflows, and the magnetic field morphologies are more structured in the polar region (see Fig. \ref{fig:2d-Bvec-B13-14-15} and Fig. \ref{fig:2d-Bvec-falloff}).

In order to estimate the total ejected mass during the simulations, we integrate the mass ejecta rate over the phase of quasi-steady state evolution. Subsequently, we multiply by the total simulation time over the time of quasi-steady state evolution, to account for the HMNS system's full evolution. We choose to integrate over the phase of quasi-steady state evolution only to exclude variable mass ejecta rate behaviour in the early stages of the simulation. The quasi-steady state phase for $\dot{M}_{\rm ejecta}$ is different for each simulation (see Panels a and b in Fig. \ref{fig:Vr-histograms-mdot}), however, in all cases, we integrate from 10 ms up to the end of the simulation. This captures most or all of the quasi-steady state phase for the majority of simulations and allows for comparison between the estimated total ejecta masses, however, for B5-15-r10 and B15-r20 the integration interval is then (partly) over a non-quasi steady state phase. We also compute the average of the mass ejecta rates over the same time interval. We list the results in Table \ref{table:massejecta} for all seven simulations. The averaged ejecta mass and mass ejecta rates for B5-15-r10 are considerably larger compared to all other simulations. This simulation, however, exhibits varying $\dot{M}_{\rm ejecta}$ behaviour throughout the evolution, meaning it does not reach a phase of quasi-steady state evolution before collapse. Despite simulations B15-r20 and B5-15-r10 both forming jets, we find much lower averaged $\dot{M}_{\rm ejecta}$ and $M_{\rm ejecta}$ values for the former, which is largely due to the $\dot{M}_{\rm ejecta}$ rapidly decreasing after $\sim 11$ ms. The averaged ejecta values in combination with the $\dot{M}_{\rm ejecta}$ evolution and larger $v^r$ velocities of unbound material for B5-15-r10 compared to B15-r20 (see Fig. \ref{fig:Vr-histograms-mdot}) indicate that a considerably more powerful jet or magnetized wind emerges in the former simulation. Except for B15-r15, all other simulations show significantly lower averaged $\dot{M}_{\rm ejecta}$ and $M_{\rm ejecta}$ compared to the jet-forming simulations. However, as we infer $M_{\rm ejecta} > 10^{-4} M_{\odot}$ for all simulations, even without jet-formation the contribution of ejected mass from the HMNS is relevant when compared to the dynamical ejecta, for which $10^{-4} M_{\odot} < M_{\rm ejecta} < 10^{-2} M_{\odot}$ has been inferred \citep{2013PhRvD..87b4001H}. Furthermore, the results in table \ref{table:massejecta} and Fig. \ref{fig:Vr-histograms-mdot} clearly show that for larger $B_0$ and $r_{\rm falloff}$, the mass ejecta and mass ejecta rates increase considerably. Similarly, the radial velocity of unbound material, shown in Fig. \ref{fig:Vr-histograms-mdot}, increases significantly for larger values of the initial magnetic field parameters of the simulations.

\begin{table}
\centering
\begin{tabular}{ | c | c | c | }
Simulation & $M_{\rm ejecta}$ [$10^{-4}\:M_{\odot}$] & $\dot{M}_{\rm ejecta}$ [$10^{-2}\:M_{\odot}$ s$^{-1}$] \\
\hline
B15-r20 & 26.8 & 12.7 \\
B14-r20 & 5.2 & 2.3 \\
B13-r20  & 2.4 & 1.1 \\
B15-r5  & 3.4 & 1.6 \\ 
B15-r10  & 6.1 & 2.8 \\
B15-r15  & 17.1 & 7.8 \\
B5-15-r10 & 83.1 & 39.8 
\end{tabular}
\caption{\label{table:massejecta} Total ejecta mass $M_{\rm ejecta}$ and averaged mass ejecta rates $\dot{M}_{\rm ejecta}$ from the HMNS outflows. For all simulations, both values are computed from 10 ms up to the end of the simulation time.}
\end{table}

In the absence of jet formation, changing the $r_{\rm falloff}$ and $B_0$ parameters of the simulations has similar effects. Namely, these simulations exhibit remarkably similar collapse times, only showing marginal differences of $\sim 0.1 - 0.2$ ms or less between simulations B13-r20, B14-r20, B15-r5, B15-r10 and B15-r15 (see Fig. \ref{fig:rhomaxvstime}). Also, they display reasonably similar mass ejecta rate evolutions (see panel a and b in Fig. \ref{fig:Vr-histograms-mdot}). Such similarities could imply small magnetic field effects on outflow properties in the absence of jet formation. However, the magnetic field parameters have considerable effects on outflow properties for other quantities, also when jets are not formed. Firstly, the radial velocities of unbound material are significantly different between the five aforementioned simulations that do not form jets (see Fig. \ref{fig:Vr-histograms-mdot}). Also, the $Y_e$ distributions and fractional abundances show apparent dissimilarities. Another indication that the magnetic fields of these simulations have considerable effects on the outflow properties is that the averaged mass ejecta rates and total ejecta mass for these simulations are significantly larger compared to the purely hydrodynamical case without magnetic field, which has been conducted by \citet{2020ApJ...901L..37M} (based on a nearly identical simulation code as this work). They find a total ejected mass of $5.8 \times 10^{-5} M_{\odot}$  and averaged mass ejecta rate of $2.4 \times 10^{-3} M_{\odot}$ s$^{-1}$ during quasi-steady state evolution. Even the lowest values for both of these quantities from table \ref{table:massejecta}, for B13-r20, are a factor $\sim 4$ and $\sim 4.5$ larger for $M_{\rm ejecta}$ and $\dot{M}_{\rm ejecta}$, respectively.

The purely hydrodynamical simulation from \citet{2020ApJ...901L..37M} does show a very similar BH collapse time of $\sim 23$ ms compared to the purely magnetized wind-forming simulations of this work. As mentioned, collapse times are partially dictated by the transport of angular momentum-carrying material out of the remnant system. It may therefore be related to the total ejecta mass during the simulation, however, this exhibits considerable differences between HD and MHD simulations as mentioned, while the collapse time does not. This implies that magnetized winds may be ineffective at transporting angular momentum out of the HMNS system compared to the mildly-relativistic jets, as the difference in collapse time between simulations with and without such magnetized winds is insignificant. Also, jet-forming simulations B15-r20 and B5-15-r10 do contain larger $M_{\rm ejecta}$ while displaying significantly smaller collapse times. These results further strengthen the case of mildly-relativistic jets being more effective at transporting angular momentum out of the remnant system compared to magnetized winds.

Increasing $B_0$ by an order of magnitude seems to have significant effects on the $Y_e$ distributions of the ejecta (when the temperature is last above 5 GK, see Fig. \ref{fig:Ye5GK}) and $r$-process yields (see Fig. \ref{fig:abundancesall}). Namely, when increasing $B_0$, the $Y_e$ distribution seems to shift to larger values while the fractional abundances exhibit lower amounts of heavy element production. Such a trend does not seem to exist for $r_{\rm falloff}$, which is especially clear when comparing the fractional abundances. However, as mentioned, this may caused by lower statistics for simulation B15-r5 (and possibly also B15-r10) due to a relatively low amount of tracer particles for this simulation, rather than being a consequence of a physical feature. 

We have shown that the strength and specific configuration of the magnetic field in post-merger magnetars can lead to robust and sizeable effects in outflow properties, such as the mass ejecta rate and radial velocity of unbound material. Indeed, in two of the seven performed simulations, the larger values of the initial magnetic field strength and falloff result in the launching of mildly-relativistic jets, thus providing characteristic electromagnetic observables. Furthermore, the change in magnetic field parameters leads to profound effects on the abundance patterns and electron fractions, and hence on the kilonova light curves. We conclude, then, that the magnetic field strength and falloff have a significant imprint on the electromagnetic observables. 

\section*{Acknowledgements}

The authors thank Luciano Rezzolla for helpful discussions regarding magnetic field configurations on merger remnants and Roland Haas for his insights regarding technical aspects of the BlueWaters supercomputer. The simulations were carried out on NCSA’s BlueWaters under allocation ILL\_baws, and TACC’s Frontera under allocation DD FTA-Moesta. The analysis of the simulations was carried out on SurfSara's Spider under the allocation EINF-2585. 

\section*{Data Availability}

Simulation data is available upon reasonable request.




\bibliographystyle{mnras}
\bibliography{sources} 







\bsp	
\label{lastpage}
\end{document}